\documentclass[smallextended,final]{svjour3}
\usepackage[utf8]{inputenc}
\usepackage[greek,english]{babel}
\usepackage{graphicx}
\usepackage{textcomp}
\usepackage{color}
\usepackage{amssymb,amsmath}
\usepackage{hyperref}
\usepackage{times}
\usepackage[authoryear,round]{natbib}
\usepackage{multirow}
\usepackage{arydshln}
\usepackage{enumitem}
\newcommand{\comment}[1]{}

\title{A not-too-simple solution to Goodman's New Riddle of induction}
\author{
  Luigi Scorzato}                                                                                  
  \institute{Accenture AG, Gen{\`e}ve, Switzerland.                                                
  \email{luigi.scorzato@accenture.com}
}
\date{}
\begin{document}

\newtheorem{defn}{Definition}
\newtheorem{ass}{Assumption}
\newtheorem{rmrk}{Remark}
\newtheorem{pstl}{Postulate}
\newenvironment{pstlp}[1]
  {\renewcommand\thepstl{#1}\pstl}
  {\endpstl}

\maketitle
\begin{abstract}
  I review the works of \citet{Gardenfors_1990} and \citet{Scorzato} and show that their
  combination provides an elegant solution of Goodman's New Riddle of induction.  The solution is
  based on two main ideas: (1) clarifying what is expected from a solution: understanding that
  philosophy of science is a science itself, with the same limitations and strengths as other
  scientific disciplines; (2) understanding that the concept of {\em complexity of a model's
    assumptions} and the concept of {\em direct measurements} must be characterized {\em
    together}.  Although both {\em measurements} and {\em complexity} have been the subject of a
  vast literature, within the philosophy of science, essentially no other attempt has been made to
  combine them.  The widespread expectation, among modern philosophers, that Goodman's New Riddle
  cannot be solved is clearly not defensible without serious exploration of such a natural
  approach.
\end{abstract}

\section{The old New Riddle}
\label{sec:intro}

Goodman's New Riddle of induction \citep{goodman1946confirmation, grue, grue83} occupies a
prominent role in the philosophy of science \citep{Henderson-SEP, Godfreysmith2, Scholz2024CC}.
In fact, it was recognized very early as deeply entangled with many other major puzzles in
philosophy of science\footnote{``{\em The urge to dispose of the problem as spurious or insoluble
  is understandable, of course, in view of the repeated failures to find a solution. The trouble
  is, though, that what confronts us is not a single isolated problem but a closely knit family of
  problems. If we set one of them aside, we usually encounter much the same difficulties when we
  try to deal with the others. And if we set aside all the problems of dispositions, possibility,
  scientific law, confirmation, and the like, we virtually abandon the philosophy of science.}''
\citep{grue83}\label{note:good-quote}}, but also one that philosophy should be capable of solving
(see in particular Putnam's preface to the 4-th edition of \citet{grue83}).  However, today, 70
years after the publication of the first edition of \citep{grue}, no solution is widely accepted
and many philosophers have no confidence that a solution is possible at all.

In this article, after a brief review of Goodman's New Riddle (Sec.~\ref{sec:gnr}), I position it
within the more general problem of scientific model selection (Sec.~\ref{sec:ts-prob}).  Then, in
Sec.~\ref{sec:perva-new}, I emphasize that both problems are very much related to other classic
problems in philosophy of science.  In Sec.~\ref{sec:what-sol}, I discuss what should count as a
solution of Goodman's New Riddle.  In Sec.~\ref{sec:new-sol}, I use the ideas of
\citet{Gardenfors_1990} and \citet{Scorzato} to formulate a simple (but not too simple) solution.
This entails a clarification of the role of direct measurements (Sec.~\ref{sec:direct-meas})
together with a well defined notion of complexity (Sec.~\ref{sec:epi-compl}).  In
Sec.~\ref{sec:just}, I argue that a justification of the solution should not fall back into a
quest to solve the old riddle (Sec.~\ref{sec:simplicity}), but it should rather show the
descriptive power of the solution (Sec.~\ref{sec:conspiracy}).  Finally, in
Sec.~\ref{sec:discuss}, I compare the proposed solution to previous proposals
(Sec.~\ref{sec:prev-sol}), focusing especially on the analysis of \citet{GardenforsStephens2017}
(in Sec.~\ref{sec:con-spa}) and the one of \citet{douglas2013value} (in Sec.~\ref{sec:values}).

\subsection{Review of Goodman's New Riddle}
\label{sec:gnr}

Goodman's New Riddle of induction \citep{grue} builds on Hume's thesis (the Old Riddle,
\citet{Hume}) that past regularities do not, by themselves, justify predictions about the future.
Goodman goes a step further because he does not question the justification of why an inductive
statement should (likely) be true about the future (he accepts that there isn't one), instead he
questions the justification of why we regard some statements as legitimate conjectures ({\em
  projectible}), while others---equally supported by the data---are not considered such.  Hence,
the New Riddle is not a puzzle about our inability to form justified expectations about the
future, which we might well have to accept. It is a puzzle about our inability to explain our own
assumptions, which is much more disturbing.

To illustrate the riddle, Goodman uses the famous ``grue'' paradox.  An object is grue if and only
if ``{\em it is observed before a specific future time $t_0$ and is green, or it is not observed
  before time $t_0$ and is blue}''.  Imagine that we have observed many emeralds and they are all
green. As long as $t_0$ lies in the future, the two sentences (a) ``All emeralds are green'' and
(b) ``All emeralds are grue'' are equally well supported by the data.  So, the question is, why do
we consider the statements (a) a legitimate extrapolation (aka {\em lawlike} or {\em
  projectible}), while (b) is not?

Before reviewing any attempt to solve the New Riddle, it is worth emphasising that the New Riddle
is about philosophy of science.  In Goodman's words: ``{\em (...) although I talk of (...) the
  color of marbles, which are seldom discussed in books on chemistry or physics, what I am saying
  falls squarely within the philosophy of science.}''  His interest are, of course, the
implications for science. Because Goodman focuses on {\bf identifying a problem}, he only needs to
discuss one example that contains the minimal ingredients necessary to highlight the specific
issue, without any distracting features.  In this sense, the sentence ``All emeralds are green''
is a stylized example of scientific model (an isolated scientific law within mineralogy).  But, if
we want to {\bf solve the problem}, we must, eventually, consider the general question of why some
models are considered legitimate scientific options, while others---equally supported by the
data---are not considered such by the scientific community.  This general question is the
well-known problem of scientific model selection, that I review in the next
Sec.~\ref{sec:ts-prob}.  Correspondingly, we can identify what Goodman calls {\em projectible}
sentences with those scientific models that the scientists often call the {\em state of the art}.
In both cases they represent those models (or sentences) that can be used for legitimate
predictions\footnote{Consistently with the {\em New} Riddle, the word {\em legitimate} here is
granted by some assumptions, that we need to identify, not by any supposed likelihood of
representing the truth.}.

Similarly, to truly solve Goodman's New Riddle, it is not sufficient to solve it in the specific
case of the grue concept. It is necessary to show that the solution is robust against any trick
that exploits similar ideas. To this end, I will introduce an extension of the idea of grue in
Sec.~\ref{sec:perva-new}.

\subsection{The problem of scientific model selection}
\label{sec:ts-prob}

How do scientists decide that some scientific models are viable options while others should be
disregarded?  {\bf Empirical evidence} is not enough. No matter how much data we have collected,
there are always infinite models that fit the data.  This observation is known as {\em
  underdetermination} of the theory\footnote{In this paper, 'theory' and 'model' are used as
synonyms, consistently with the common informal practice, among scientists, of referring to
'models' as simple/unambitious theories. Model theory, in the sense of mathematical logic, does
not play a significant role in this article.} by the data \citep{Duhem1954, quine1975empirically,
  sep-scientific-underdetermination}.

These infinite options are not just theoretical possibilities without practical relevance.
Consider statements like: {\em ``was my experimental device malfunctioning on day
  X?''}\footnote{We can test some of these statements, but only a tiny fraction of the possible.}.
This is the very concrete way in which infinite options of ad-hoc assumptions\footnote{Often
scientists argue that these are not scientific models, but a definition of {\em scientific model}
that excludes these options without excluding many other legitimate options is not available.} can
be exploited to fit any data.  This kind of questions emerge very often in real scientific
practice, and they can be very challenging to resolve.  However, they do not seem to pose an {\em
  insurmountable} obstacle to the scientific practice.  Which other tools do the scientists use to
discriminate among these equally accurate infinite options?

A good observation is that state-of-the-art scientific theories are typically {\bf
  predictive}. However, predictions are also not enough.  A successful experiment might increase
the probability of a model, in some sense.  \citet{Carnap-conf} devoted enormous efforts to this
problem (see \citet{sep-confirmation} for a review and \citet{Leitgeb2024-VC} for a recent
contribution on this topic).  But can we translate these results into a rule for model selection,
even a very crude and approximate one?  Unfortunately, we cannot.

The reason is the same behind the impossibility of determining the ``right'' {\em p-value} (even
approximatively) that can be used to confirm a discovery \citep{goodman2008pvalue}.  In fact, such
hypothetical {\em p-value} should be extremely small (say, at least $p \ll 10^{-10}$) to prevent
us from rejecting our best scientific theories in favor of a crazy alternative that happens to
correctly guess some very unlikely events.  On the other hand, the same {\em p-value} should be
$p \gtrsim 1/3$ to justify a vast amount of valuable model selections that enjoy the unshakable
support of the scientific community (especially in domains where impressive predictions are rare).

Probability alone cannot determine model selection.  Predictions are valuable only if they are
based on {\em good} assumptions.  We obviously need to assume also some {\bf non-empirical}
(i.e. non evidence based) {\bf epistemic value} to justify the scientific model selections that
are regularly adopted by the scientific community.  Indeed, many scientists and philosophers have
recognized that some non-empirical epistemic values play a formal role in model selection.
Einstein, for example, famously said \citep{EinsteinLife}: ``{\em The grand aim of all science is
  to cover the greatest number of empirical facts by logical deduction from the smallest number of
  hypotheses}''.  Many others scientists and philosophers made similar statements.

\subsection{The New Riddle of induction and its pervasiveness}
\label{sec:perva-new}

Unfortunately, Einstein's characterization of the goal of science has a major flaw: how do we {\em
  count} the number of hypotheses?  In fact, one can always introduce a new symbol $\Xi$ to
express all her hypotheses as $\Xi = 0$, and the number of hypotheses would be just one!  This
fact was noted by many \citep{Feynman1963, Kelly-efficiency, Votsis2016, Scorzato}, and it is
reviewed here in Appendix \ref{sec:app_xi}.

As noted in Sec.~\ref{sec:gnr}, the problem of scientific model selection can be seen as a
generalization of the New Riddle of induction. Along the same lines, the $\Xi$ trick mentioned
above can be seen as an extension of the idea behind Goodman's grue, and it emphasises the same
lesson: for every crazy theory that agrees with the data, there is always an ad-hoc language that
makes it a simple and/or natural assumption. The $\Xi$ trick simply emphasises how general and
deep the New Riddle is.  The notion of simplicity seems to be irremediably subjective, hence
unsuitable to define the goals of science.

It is important to understand that the $\Xi$ trap does undermine only syntactic simplicity.
Although the need for a non-empirical epistemic value is clear, no one seems immune to the $\Xi$
trap.  For example, according to \citet{Baker-SEP}, a theory is more {\em ontological
  parsimonious} than another if it postulates less entity types.  However, nothing prevents us to
introduce a new entity type that stands for the union of all the types of the theory, and the new
formulation has just one type.  As another example, a long philosophical tradition, going back at
least to \citet{Popper} and recently reviewed by \citet{schindler2018theoretical}, assigns special
importance to whether a theory makes ad-hoc assumptions and how many.  But if the only hypothesis
of the theory is $\Xi=0$, then the distinction between ad-hoc and non ad-hoc hypotheses cannot be
defined.  I review more candidate epistemic values in Sec.~\ref{sec:values}.

Goodman's New Riddle does not only hinder a characterization of induction and theory
selection\footnote{Many of the problems discussed in this section come in two
versions, depending on whether they aim at justifying one choice in terms of likelihood of future
success (e.g. the Old Riddle), or they simply aim at describing the scientists' actual choices
(e.g. the New Riddle).}.  As emphasised by his quote reported in Footnote \ref{note:good-quote},
the puzzle permeates much of philosophy of science.  Goodman himself discussed in detail how the
New Riddle is essentially equivalent to the problem of {\em counterfactuals}
\citep{sep-counterfactuals}.  In fact, deciding what would happen in a hypothetical, unrealized,
scenario corresponds to selecting a class of models that are admissible to draw those conclusions.
If any model that fits the empirical data is legitimate, almost any conclusion is possible, and
counterfactual reasoning loses any meaning.  A vast literature \citep{sep-dispositions} also
connects the problem of counterfactuals to the problem of {\em dispositions}.

The problem of induction (and, therefore, the problem of model selection) is also strongly
entangled with the problem of {\em confirmation} \citep{sep-confirmation}.  In fact, both aim at
identifying which models should be used for plausible predictions and which shouldn't. A solution
to one problem would yield a solution to the other, but no solution enjoys widespread support.  In
particular, Bayesian confirmation theory \citep{Bayesian, SprengerHartmann-bayesian} had cherished
the hope of establishing confirmation on empirical evidence and probability alone.  However, also
Bayesianism must rely on something more to be able to deliver any meaningful outcome
\citep{sep-science-theory-observation, norton2021material}. See also Sec.~\ref{sec:prev-sol}.

The classic problem of {\em demarcation} \citep{demarcation-IEP} also depends on the problem of
model selection.  The former is often seen as the problem of excluding entire disciplines as
non-scientific at all, rather than individual models.  But, if we cannot rule out a model as
implausible as ``All emeralds are grue'', it is obviously also hard to rule out, conclusively, any
pseudo-scientific theory (and even less a discipline), provided that it has done its job of
ensuring concordance with the empirical evidence, which is not too difficult, by introducing a
sufficient amount of ad-hoc hypotheses.

Furthermore, any idea of {\em scientific progress} that goes beyond a dull list of empirical
results, must involve the idea that our set of state-of-the-art theories is somehow improving.  It
is difficult to imagine how this can be achieved without a definition of state-of-the-art or a
definition of theory selection.

Finally, the analysis by \citet{roche2023purely} shows that no existing, purely probabilistic,
measure of {\em explanatory power} is satisfactory.  On the other hand, one of the most popular
account of explanation \citep{woodward2005making} relies crucially on counterfactual reasoning,
which, as we have seen above, also relies on a solution of the New Riddle.  Another, very
influential account of explanation \citep{Kitcher1989} relies on recurring patterns in scientific
laws, an argument that loses any differentiating power, if any theory is expressed as $\Xi=0$.

These deep interdependencies between some of the most fundamental philosophical issues are not
surprising.  The $\Xi$ trick offers an interesting perspective into the core problem: if you can
write all your laws as $\Xi = 0$ (and you certainly can), and you can measure $\Xi$ directly
(which is the weak point, as we will see, but difficult to contest, as there is certainly a well
defined correspondence with what we do measure), and if the law $\Xi = 0$ is accurate (which is
true by construction), how can this theory be less than optimal by any standard?  It is very hard
to imagine any well-defined and meaningful comparison between different---equally
accurate---models, if they can all be represented in this way. But such comparison is a
precondition to address any of the philosophical problems mentioned above.

In conclusion, Goodman's New Riddle and its extensions seem to undermine any precise definition of
{\em theory selection, goals of science, induction, demarcation, confirmation, scientific progress
  or explanation}.  Indeed, no definition of any of these concepts has gained a consensual
endorsement of the philosophical community and the only definitions that gained some popularity
inevitably refer (more or less explicitly) to the unconstrained judgment of the scientific
community.  This necessarily falls short of identifying the hidden assumptions behind the
scientists' decisions. It also doesn't explain why, 60 years after Kuhn explained to the
scientists that {\em they} are the ultimate source of truth of theory selection, they still often
refer to their conclusions as 'objective' and 'fact-based'.  Finally, this framework cannot help
decide any controversial cases, which makes philosophy of science hardly useful, especially in a
time when it would be most needed, because AI significantly blurrs the distinction between
scientists and non-scientists.

Until at least the 80's, prominent philosophers held the firm belief that philosophy should be
able to offer a satisfactory solution to Goodman's New Riddle (see e.g. the preface of
\citet{grue83}).  However, in the past few decades, the philosophical community has increasingly
accepted the idea that also the New Riddle, like the Old one, might remain insurmountable.  But
Goodman's intuition remains valid: as I have just reviewed, the scientists obviously use some
unspecified assumptions in their decisions of model selection.  Then, there are only two
possibilities, either (a) we can identify such assumptions, which is the only way to, perhaps,
judge whether they are acceptable; or (b) we cannot.  But option (b) implies that any scientific
conclusion relies, inexorably, on fundamentally mysterious assumptions.  But an essential step in
any scientific work consists in identifying every hidden assumption in the arguments used by
scientists.  Any effort in this sense becomes futile, in the scenario (b).  It does not help to
formulate scenario (b) with narratives that obfuscate how much it fundamentally undermines the
value of science.  Fortunately, scenario (b) is just not plausible.  But then we must first
understand what is wrong with the formulation $\Xi = 0$.  This must be possible: {\em If there
  aren't objective standards, [we must] construct standards!}  \citep{grue83}.

\subsection{What does count as a solution?}
\label{sec:what-sol}

A major obstacle in solving Goodman's New Riddle lies already in the confusion of what should
count as a solution \citep{Scholz2024CC}.  In particular, requiring that a candidate solution is
justified in terms of likeliness of future successes is not legitimate, because it represents a
relapse in the quest for a solution to the {\em Old} Riddle, that we cannot expect to solve.  On
the other hand, a mere enumeration of the scientists' actual choices is not satisfactory either:
it would be useless to interpret any new model selection.

An acceptable solution should describe all cases of model selection supported by broad scientific
consensus, but it should do it by providing a {\em general rule}\footnote{See in particular
\citep{grue83}: {\em``what we want, indeed, is an accurate and general way of saying which
  hypotheses are confirmed by (...) any given evidence''}.}  and not a mere list.  What is a
general rule?  It is a concise rule that covers most cases with few or no exceptions.  But, as we
just saw in the case of scientific laws, it is always possible to forge a general rule from a mere
list of examples by using a $\Xi$ trick.  This makes it clear that {\bf deciding what counts as a
  solution to Goodman's New Riddle} is equivalent to {\bf solving Goodman's New Riddle in the
  special case when the scientific model under scrutiny is a model for scientific model selection
  itself}.  This is not very surprising, if we accept that philosophy of science is a science
itself, it should demand of itself what it demands of other scientific disciplines.  This provides
further evidence that little makes sense in philosophy of science unless we understand what is
wrong, exactly, with the formulation $\Xi = 0$.  But everything changes once we can clarify that
conundrum.  Once this is done, I come back to the question of what counts as a solution in
Sec. ~\ref{sec:just}.

\section{A simple, but not too simple, solution of the New Riddle}
\label{sec:new-sol}

To understand what is wrong with $\Xi = 0$, it is not enough to argue that $\Xi$ is perhaps hard
to measure, if at all.  We need a clear criterion that excludes the formulations that should be
excluded, but not more.  The key observation is that, although we can express any model as $\Xi =
0$, we cannot, at the same time, expect measurements in the form of a central value and a convex
error-bar as $\Xi = \Xi_0 \pm \Delta $.  In fact, the general argument leading to a reformulation
in the form $\Xi = 0$ (see Appendix \ref{sec:app_xi}) does not extend to an argument guaranteeing
the preservation of the convexity of the error-bars.  Further insight is offered by the following
heuristic argument: if it were possible to express any measurement as $\Xi = \Xi_0 \pm \Delta $,
we would know from the formulation of the model itself what is measurable and what is not and the
expected precision of any measurement.  But we do not have this information for most real modern
theories.  So, this representation cannot be logically and empirically equivalent to any realistic
modern theory.  Chaotic billiards \citep{Scorzato}, deep neural networks
\citep{scorzato2024reliability} and the measurement of grue at $t_0$ (later in this section)
provide examples where the limitation are implied by current physical laws. The measurement of
grue at time $t \gg t_0$ and the example in Sec.~\ref{sec:conspiracy} show that the limitation may
also be due to practical reasons.

Can we use this idea to define model selection? In fact, we can. I sketch here briefly the key
ideas that I will develop in the rest of this section.  We are looking for a notion of complexity
of the assumptions (epistemic complexity, Def.~\ref{def:complexity}) that can be combined with
accuracy to determine model selection (Def.~\ref{def:selection}).  To be plausible, such notion
should be invariant by reformulation of the model.  Invariance can be easily achieved by taking
the minimum over all possible equivalent reformulations.  But we have seen that, if we consider
all {\em logically equivalent} reformulations, the minimum is trivial.  The key is to realize that
logical equivalence is not sufficiently restrictive.

In fact, two formulations are truly equivalent only if they are also {\em empirically equivalent}
(Def.~\ref{def:eq}), which doesn't follow automatically from being logically equivalent.  Two
formulations are empirically equivalent if they predict measurement outcomes that are consistent
with the same precision in both formulations.  When we formulate the assumptions of a model, we
must include a list of the quantities that we assume to be directly measurable.  We do not need to
include anything that we can measure, but at least enumerate a basis that is sufficient to define,
operationally, all the other measurable quantities\footnote{Scientific models often include also
theoretical concepts, which cannot be operationally defined from directly measurable quantities.}.
Such basis must be part of the model assumptions to claim unambiguous interpretation of the
model's empirical content (Def.~\ref{def:model}).  There is much freedom in the choice of the
basis, but it is also constrained by the need of (i) enabling an operative definition of any other
measurements and (ii) being plausibly directly measurable, which entails the minimal requirement
identified in Post.~\ref{def:P1}.  This is where convexity plays a crucial role.  These
constraints are now sufficient to ensure that the shortest formulation among all logically {\em
  and empirically} equivalent ones, is, in general, not trivial anymore.

\subsection{A postulate on direct measurements}
\label{sec:direct-meas}

A necessary requirement of any direct measurement can be identified in the following:

\begin{pstl}
  \label{def:P1}
  The result of a valid single direct measurement of ($k$-dimensional) property $Q$ is always
  expressed as a ($k$-dimensional) central value $Q_0$ and a ($k$-dimensional) convex set
  (error-box) that contains $Q_0$.
\end{pstl}

\begin{figure}
\centering
\includegraphics[height=.3\linewidth]{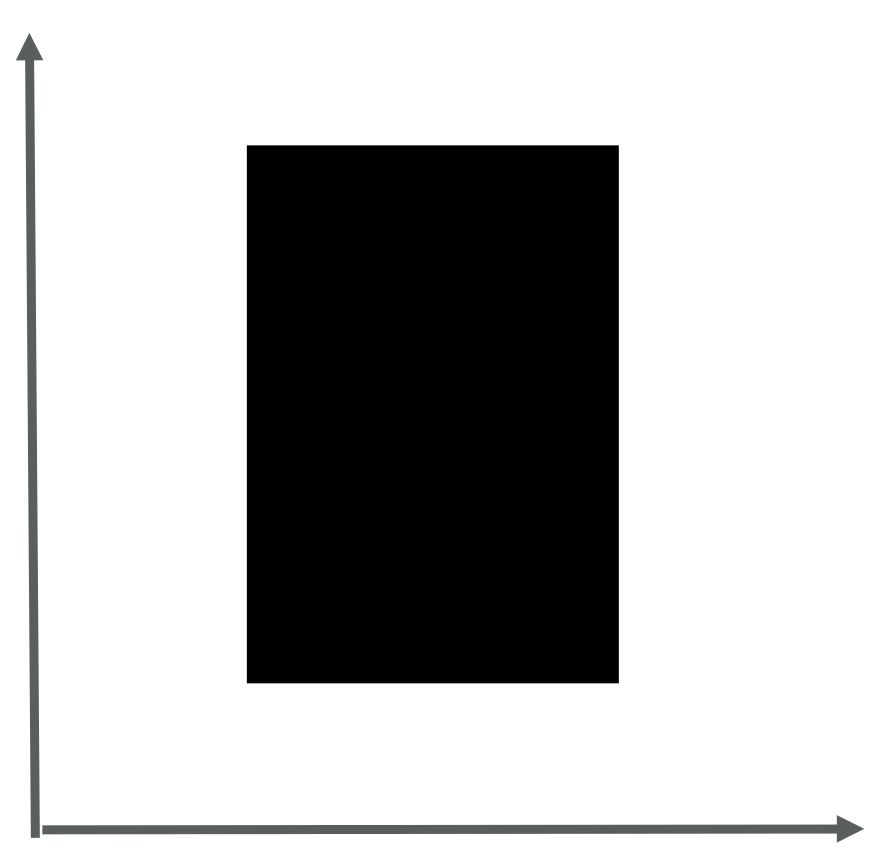}
\hspace{2cm}
\includegraphics[height=.3\linewidth]{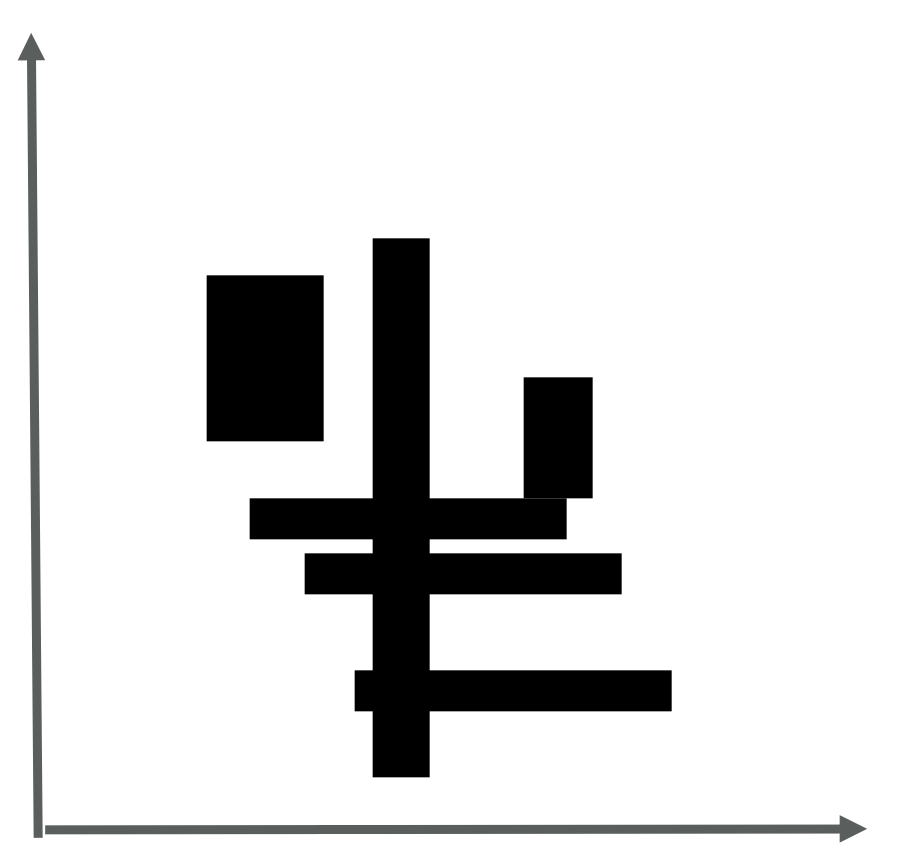}
\caption{\label{fig:errorbox}Left: a legitimate error-box outcome of a single direct measurement in
  two dimensions. Right: not a legitimate one.}
\end{figure}

In this section I justify why we are allowed to make this general assumption and why it is useful
to do so.  Post.~\ref{def:P1} reflects the scientific practice of quoting error-boxes as a
$k$-cell\footnote{A $k$-cell is a $k$-dimensional box.  In the context of error-boxes, the
difference between a convex set and a $k$-cell is not significant.  In fact, any $k$-dimensional
box is convex and in any convex set we can inscribe an $k$-dimensional box
\citep{behroozi2022rectangleinconvex}.  For simplicity, we use the name error-box in the general
case.}.  For example, if we measured the temperature of a room once, it makes no sense to say that
the result was ``either $20 \pm 1 C^{\circ}$ or $30 \pm 1 C^{\circ}$''.  This outcome might
potentially result from multiple direct measurements (e.g. taken under two types of conditions) or
from a derived measurement (e.g. obtained as the solutions of the constraints from other
measurements), but not from a single direct one.  Another example is shown in
Fig.~\ref{fig:errorbox}.  The error-box on the left is a legitimate outcome a single direct
measurement in two dimensions.  On the contrary, the black region displayed on the right cannot
represent a legitimate outcome of a single direct measurement.  Note that I have not defined {\em
  direct measurement} explicitly: it is up to the scientific theory to select the most appropriate
quantities, but, whatever they are, they must fulfill Post.~\ref{def:P1}.  More on this after
Def.~\ref{def:model}.

\paragraph{An extension of Post.~\ref{def:P1}.}
Although Post.~\ref{def:P1} is very intuitive, we should remark that error-boxes are merely
short-hand notations for features of the expected probability distribution of a measurement.  It
is therefore worth introducing an alternative formulation in term of probability distributions, as
in the following:

\begin{pstlp}{1'}
  \label{def:P2}
  The contour sets\footnote{The contour set of $P$ at level $l$ is the set $\{Q:P(Q)\geq l \}$.}
  of the expected probability distribution $P(Q)$ of a valid single direct measurement of a
  property $Q$ are convex sets, for each level $l$.
\end{pstlp}

\begin{figure}
\centering \includegraphics[height=.2\linewidth]{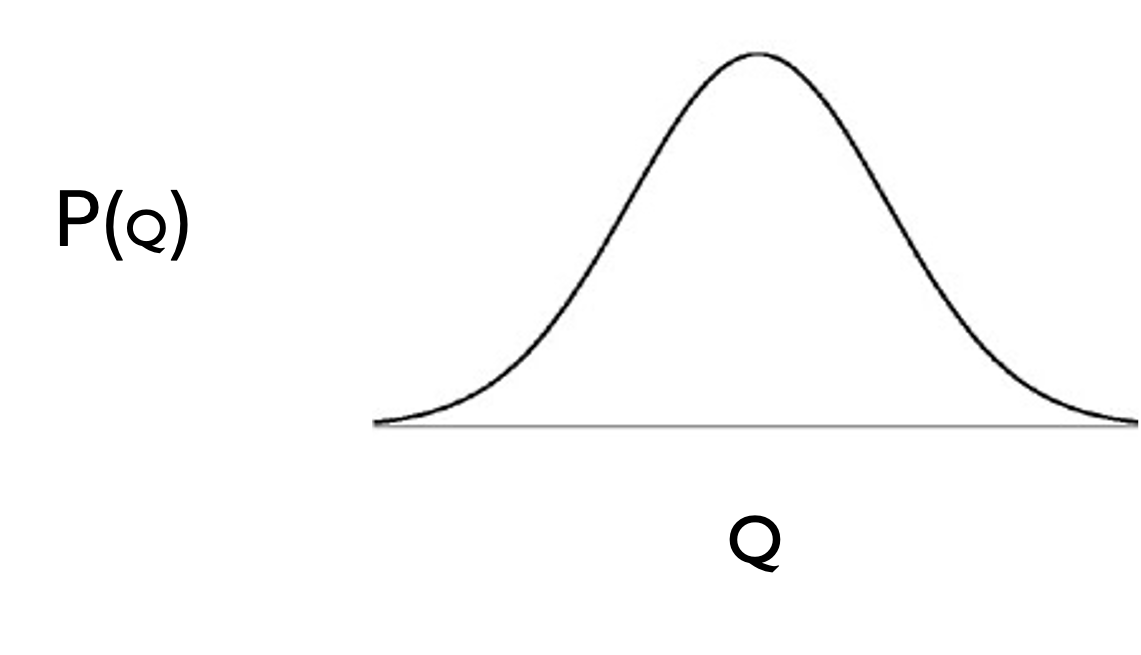}
\hspace{2cm}
\includegraphics[height=.2\linewidth]{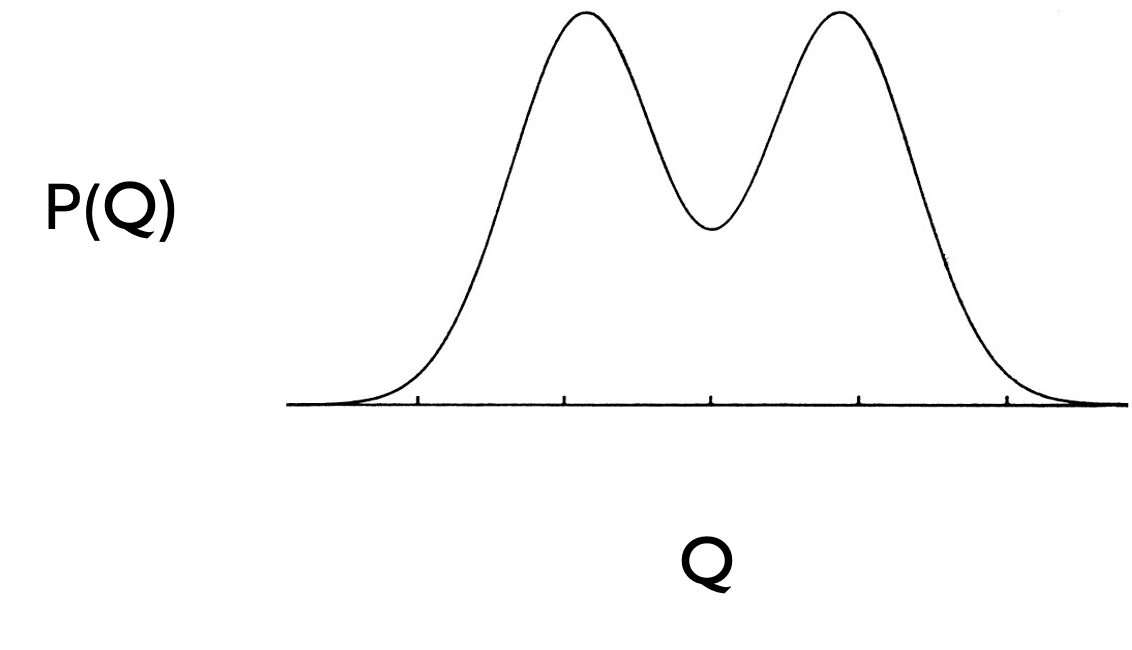}
\caption{\label{fig:prob}Left: a legitimate probability distribution of a single direct
  measurement in one dimension. Right: not a legitimate one.  Note that this is {\em not} the {\em
    empirical} distribution of {\em multiple} measurements, but the {\em expected} distribution of
  a {\em single} measurement.}
\end{figure}

In particular, the distribution on the left of Fig.~\ref{fig:prob} is a legitimate expected
distribution of a single (one dimensional) measurement, while the distribution on the right of
Fig.~\ref{fig:prob} is not\footnote{It is not difficult to extend Post.~\ref{def:P2} to measurable
quantities defined on discrete sets.  This is done in Appendix \ref{sec:app_discrete}.}.  It is
important to emphasise that $P$ is {\em not} the {\em empirical} distribution of {\em multiple}
measurements, but the {\em theoretical expected} distribution of a {\em single} measurement.  The
former could very well be multimodal, but the latter cannot\footnote{Note that even the
double-slit experiment, that plays an key role in Quantum Mechanics (QM), does not contradict the
claim. What is measured directly is the interaction of single electrons with the photographic
plate. These individual dots have a convex form. Even if the QM explanation of the double-slit
experiment implies that the wave function of a single electron traverses both slits at the same
time, the wave function itself is not measured at all (not even indirectly), and the interference
pattern is only visible after multiple hits.}.  In the following, I will refer to
Post.~\ref{def:P1}, for simplicity.  But one can easily (although tediously) verify that all
important conclusions would be maintained under the more general Post.~\ref{def:P2}.

\paragraph{Relation to G\"{a}rdenfors' convexity.}
Post.~\ref{def:P1} represents a special case of the requirement proposed by G\"{a}rdenfors that
natural properties form convex sets in conceptual spaces \citep{Gardenfors_1990, Gardenfors_2000,
  GardenforsStephens2017}.  A deeper comparison with G\"{a}rdenfors' approach is discussed in
Sec.~\ref{sec:con-spa}.  Here, I only note that I do not introduce the concept of {\em natural
  properties}.  Instead, I require Post.~\ref{def:P1} only for directly measurable properties and
only for the small regions that correspond to the uncertainty of a measurement\footnote{A related
idea has been proposed also by \citet{schurz2015ostensive}, because {\em ostensively learnable}
concepts must be measurable and hence convex, according to Post.~\ref{def:P1}.  However,
\citet{schurz2015ostensive} characterizes ostensive learnability only empirically, which is not
sufficient to identify the implicit assumptions in model selection.}.

\begin{figure}
  \centering
  \includegraphics[height=.2\linewidth]{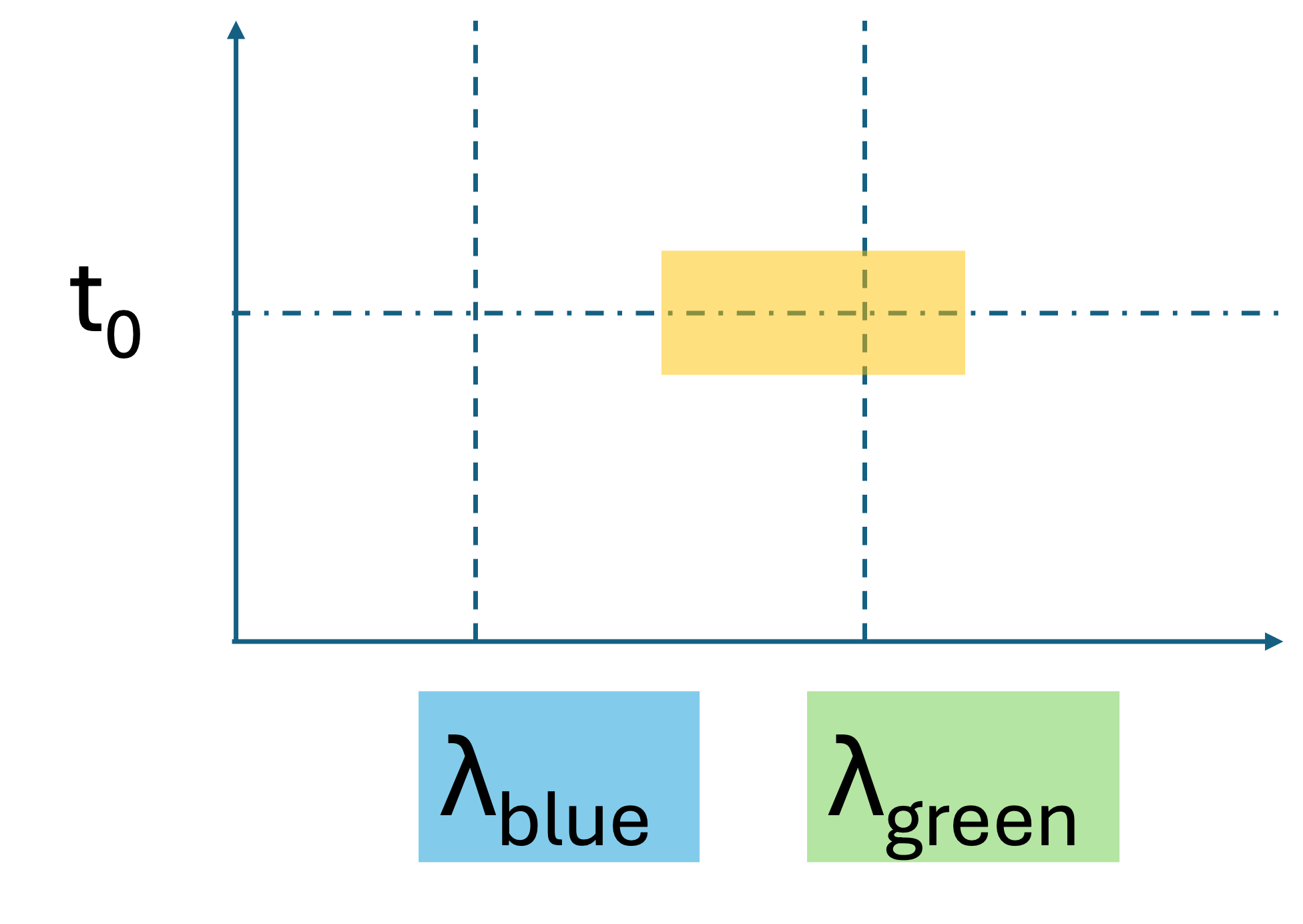}
  \hspace{2cm}
  \includegraphics[height=.2\linewidth]{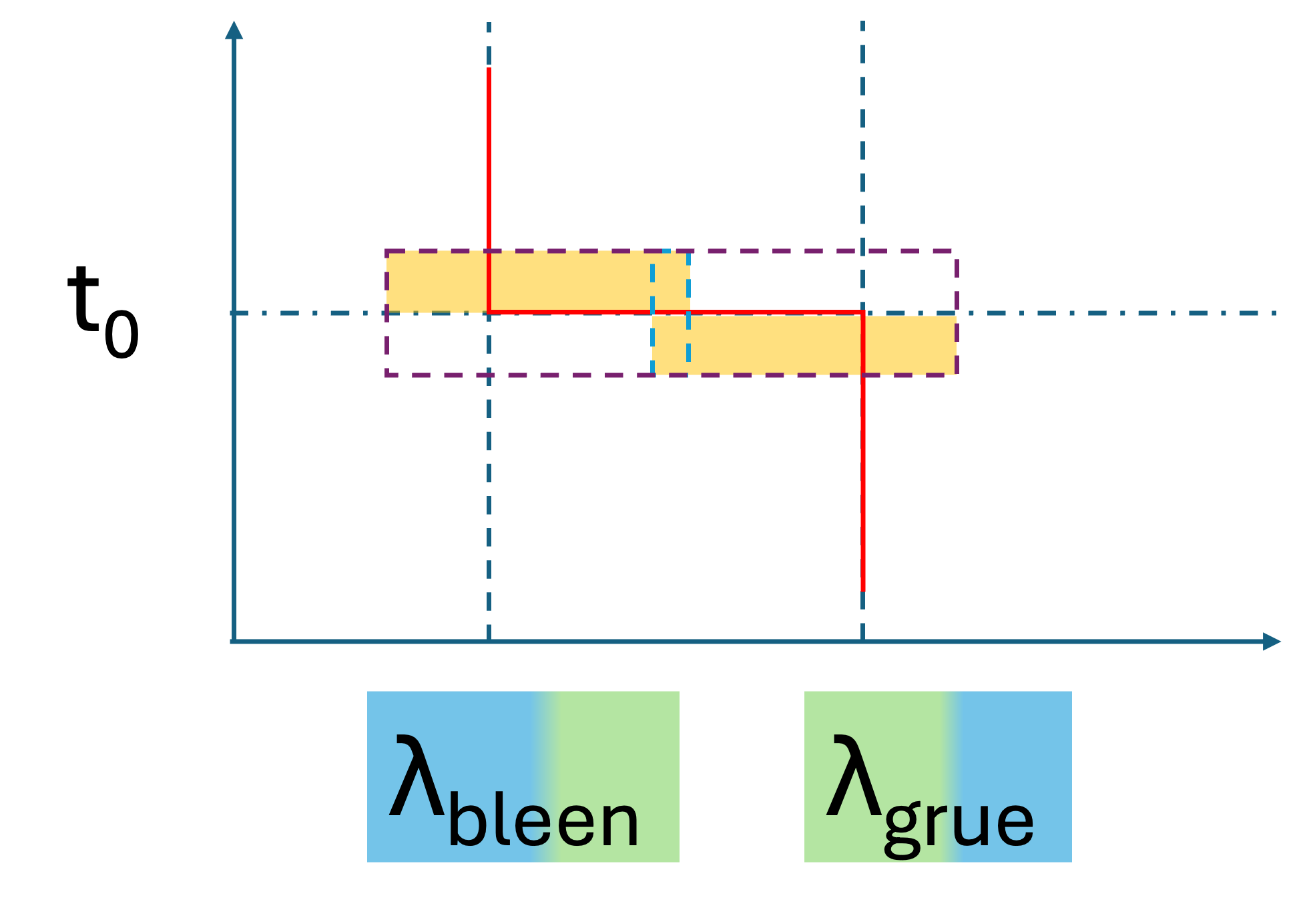}
  \caption{\label{fig:grue}Left: error-box (in yellow) associated to the measurement of the color
    of an emerald around the critical time $t_0$. Right: the same error-box appears split in the
    grue/bleen representation. Cfr Fig.~4 in \citep{Gardenfors_1990}.  On the right side, one can
    enforce a convex error-box only by either admitting more possibilities than the measurement
    attest (the large dashed-violet box), or by admitting less (the small dashed-cyan box).  Both
    options end up reducing the accuracy of the model. The plot on the right assumes an overlap
    between the error-bar around green and the one around blue, which is typical of low precision
    measurements. The overlap disappear for higher precision measurements. This is only meant to
    stress that connectivity of the error-boxes is not the relevant property: convexity is.}
\end{figure}

\paragraph{Implications for grue.}
What does the previous discussion imply for Goodman's grue?\footnote{Note that in this paper, I
adopt the definition of grue originally proposed in \citep{grue}: emeralds are grue iff they are
green and they are {\em first seen} before $t_0$ or they are blue and they are {\em first seen}
after $t_0$. So, individual stones do not change color.  We could alternatively define grue
emeralds iff they are green before $t_0$ and blue after $t_0$ \citep{Gardenfors_1990}.  In the
latter model, individual stones would change color at $t_0$.  As correctly noted by
\citet{Gardenfors_1990}, the core idea of Goodman's New Riddle and G\"ardenfors' argument remain
unchanged.  In both cases, there is an uncertainty on the measurement of time, whether it is the
time of first observation or the time of a subsequent observation, and an uncertainty on the
measurement of wavelength.  However, if we use G\"ardenfors' model, we must distinguish two cases.
If the stone does not change color at $t_0$, Fig.~\ref{fig:grue} remains unchanged, both right and
left. If the stone is observed while it does change color (at around $t_0$), the horizontal
error-bar necessarily increases (both right and left). In this case, the error-box doesn't
completely split, but the measurement is less precise.  This does not change the conclusion about
non-convexity and lack of direct measurability of grue.

On the other hand, Goodman's original definition has an additional interesting twist for times $t
\gg t_0$, because it is impossible to determine the grue/bleen property of an emerald at later
time, unless we keep track of when each emerald was first observed.  G\"ardenfor's definition does
not have the same problem for $t \gg t_0$.}  If I measure the color of an emerald at the critical
time $t_0$ (when the color of newly seen emeralds might be blue, according to Goodman's model), I
have two uncertainties: one on the color wavelength and one on time (see left panel of
Fig.~\ref{fig:grue}).  If I translate this observation from the blue/green representation into the
grue/bleen one, I cannot be sure if the emerald is seen before or after $t_0$, so the error-box
gets split, consistent with the fact that I measured directly color and time, not grue/bleen
colors (see right panel of Fig.~\ref{fig:grue}).

Grue {\em is} measurable in the sense that I can estimate its value (I just need a colorimeter and
a clock).  A jump on its value does not undermine measurability: many physical quantities display
jumps.  But when I measure directly a quantity with jumps, I see a {\em large} error-box, not a
{\em split} error-box.  Split error-boxes are incompatible with a {\em direct} measurement,
although they are fully acceptable for indirect measurements.

\paragraph{A common objection.}
A common objection is the following: {\em why couldn't we imagine an experimental device that
  measures grue directly and hence produces convex error-boxes?}  This objection tries to
replicate the symmetrization trick that Goodman himself used when arguing that we could regard
grue and bleen as fundamental, rather than green and blue.  But, here we face a different
obstacle: while we have much freedom to decide what we consider logically fundamental, we do not
have the same freedom to decide what we can measure directly.

It is instructive to explore in detail how one might try to achieve convex error-boxes for grue.
One can either (a) try to design an experimental device that measures grue directly, or (b) {\em
  pretend} of doing it.  Let us consider (b) first.  I emphasise that I never assume that any
proponent of an alternative model is honest: she could secretly take measurements of ordinary
colors and time, and then convert them into measurements of grue, pretending that they had
originally been recorded directly in that form.  Because the original measurements were actually
taken by combining color and time measurements, the error-boxes become non-convex in the grue
variable, as in the right panel of Fig.~\ref{fig:grue}.  The error-boxes could be fraudulently
modified, but this inevitably compromises the accuracy of the model.  In particular, enforcing
convexity using larger error-boxes (e.g. the large dashed-violet box in the picture) would assign
a non-zero probability to scenarios (the off-diagonal sub-boxes) excluded by the original
measurement, thereby reducing the model's accuracy.  Alternatively, enforcing convexity with
narrower error-boxes (e.g. the small dashed-cyan box in the picture) would claim a level of
precision higher than that actually granted by the device; this would, once again, lead to
discrepancies in certain scenarios and compromise the model's accuracy.

In case (a), we can easily build a native grue detector if we accept larger error-boxes than those
of the standard model, as the dashed-violet box in the right panel of Fig.~\ref{fig:grue}.  But if
we need to ensure that the accuracy of the grue/bleen model remains competitive with the standard
model, this approach is not sufficient.  The only option is to postulate new electrodynamics laws
and new directly observable quantities that might achieve both convexity and precision.  Without
investigating how these new laws might look like, we know that they must postulate at least one
new observable quantity, which must be independent from the existing ones, because its convexity
properties are incompatible with the existing ones.  That means that the grue/bleen model requires
additional directly measurable quantities.  These cannot just replace the existing directly
measurable quantities because, even in the grue/bleen model, there are objects that remain green
across $t_0$ and the grue/bleen model is supposed to describe the related phenomena as accurately
as the standard model\footnote{Note that if every green object becomes blue and every blue object
becomes green at $t_0$, then the grue/bleen model is just a trivial renaming convenction, not an
alternative model within mineralogy, as was the intention of Goodman.}.  Nor can we pretend that
these different directly measurable quantities are just one and the same (i.e. apply again a $\Xi$
trick), because such combined measurements would break convexity.  This is how convexity forces
the introduction of {\em additional symbols} for the grue/bleen model.

In conclusion, we can maintain that we measure grue directly only at the expenses of either
accuracy or the introduction of additional symbols.  In the first case, the grue/bleen model is no
longer empirically equivalent to the green/blue one.  In the second case, it is strictly more
complex, in the sense that I will make precise in Sec.~\ref{sec:epi-compl}. Hence, the convexity
of error-boxes highlights a fundamental, well-defined, asymmetry between the green/blue and the
grue/bleen models.

\begin{figure}
  \centering
  \includegraphics[height=.2\linewidth]{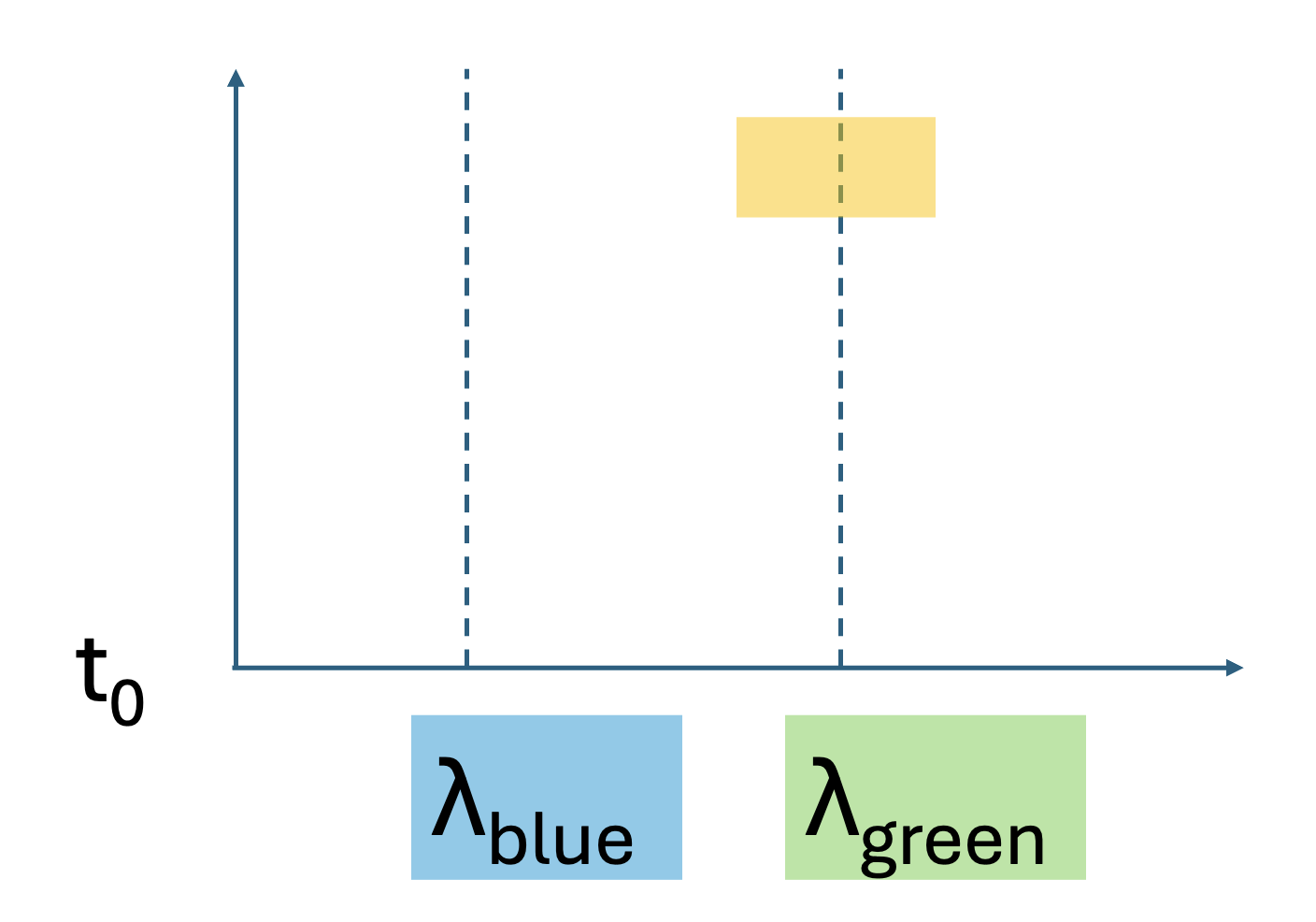}
  \hspace{2cm}
  \includegraphics[height=.2\linewidth]{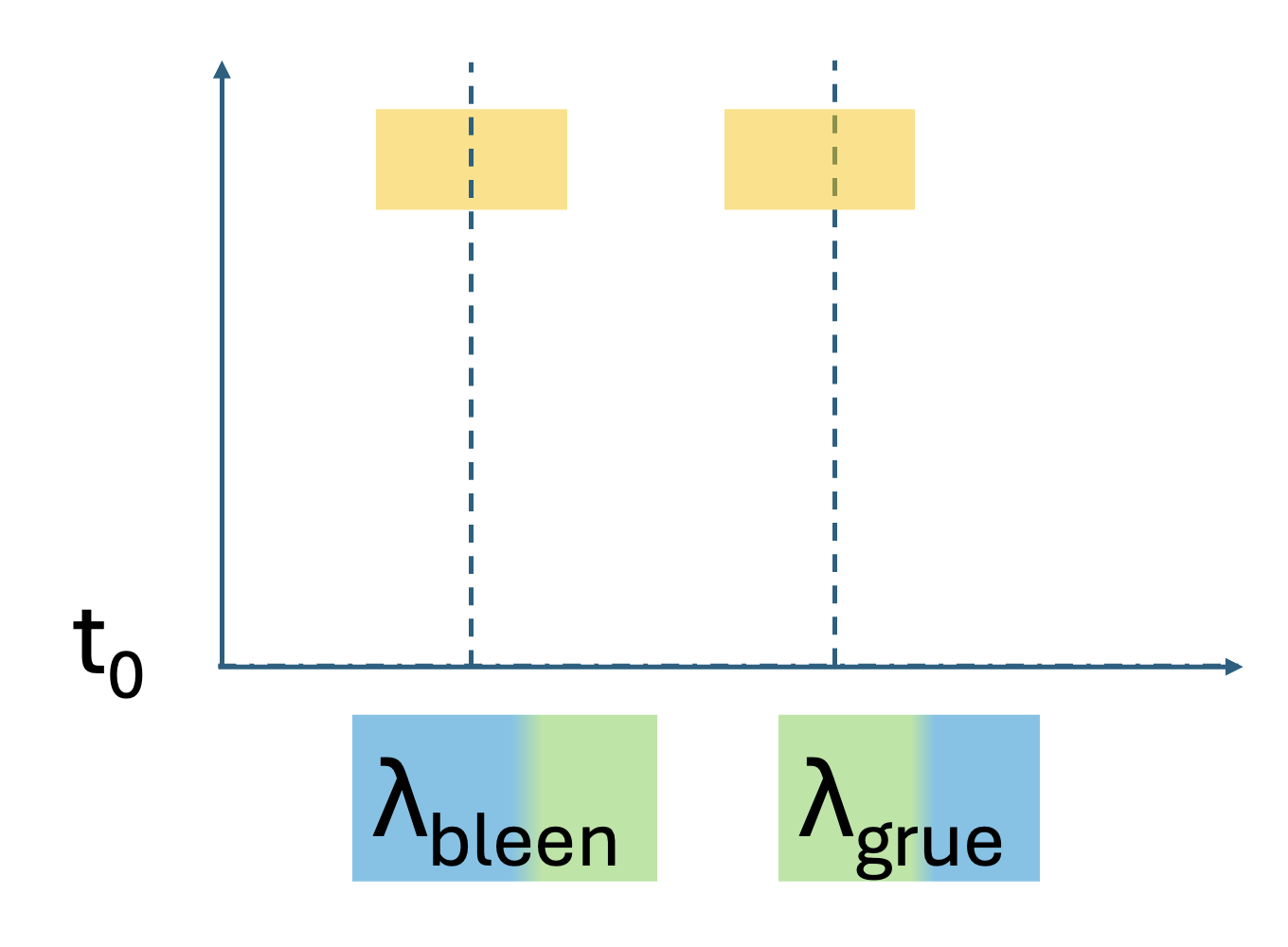}
  \caption{\label{fig:grue_aftert0}Left: error-box (in yellow) associated to the measurement of
    the classical color of an emerald well after the critical time $t_0$. Right: the same
    error-box gets split in two disconnected boxes in the grue/bleen representation, if we
    cannot tell whether the emerald was first seen before or after $t_0$.}
\end{figure}

\paragraph{A less fundamental, but more common, failure mode for grue.}
Besides the violation of convexity around time $t_0$ discussed above, there are more problems even
well after $t_0$.  To appreciate it, consider a bunch of green emeralds at time $t \gg t_0$.  Are
they grue or bleen? They would be grue if they were first seen before $t_0$, and bleen if they
were first seen after $t_0$.  But, if the time of first detection {\em has not been recorded},
then we cannot tell: they are undetermined in the grue/bleen dimension.  In other words, for the
set of emeralds that we can practically collect, the grue property is not convex\footnote{Here I
rely on \citet{Gardenfors_1990} extension of the notion of convexity to discrete sets. See also
the Appendix \ref{sec:app_discrete}.\label{fn:convex}}, even well after $t_0$, as depicted in
Fig.~\ref{fig:grue_aftert0}.  This example highlights a limitation of grue that is not traced back
to our curent physical laws---like the impossibility to measure time with arbitrary high
precision---but it is nevertheless a practical limitation, because we do not record all possible
information (even when we could), if we do not consider it useful.  This scenario is very
important in practice, as I will discuss in Sec.~\ref{sec:conspiracy}, and it is reminiscent of
Goodman's own solution that relies on {\em entrenched} concepts.  However, the logic is very
different from the one proposed by Goodman, as I will clarify in Sec.~\ref{sec:prev-sol}.

\vskip 0.2 cm

\paragraph{Not yet a solution.}
Does the requirement of convexity solve Goodman's New Riddle?  More explicitely, shall we state
that projectible models are those expressed in terms of convex quantities?  Such requirement would
be both too weak and too strong. It is too strong because scientific theories rely fundamentally
on concepts that are not even measurable (e.g. the quark wave function).  For non-measurable
concepts, the choice of the metric that determines convexity is too arbitrary to offer a useful
selection.  On the other hand, the requirement is also too weak because by using only
convex properties we can still build accurate but implausible models.  For example, consider a
model based on a simplistic general rule together with a long list of exceptions.

However, the idea of convexity does achieve an important result: it represents a well defined,
general and detectable property that breaks the symmetry between green and grue.  As a result,
defining the grue/bleen model for emeralds is more {\em complex} than defining the standard model.
In fact, if we insist that any model must introduce, as part of its definition, all directly
measurable properties that are necessary to derive its conclusions, then the concept of grue has
an objective disadvantage with respect to the concept of green, because it needs an {\em
  additional} step to be defined.  But, to make this statement precise, we now need to define a
notion of complexity of (the assumptions of) a model.  This is the goal of the next section.

\subsection{Epistemic complexity and scientific model selection}
\label{sec:epi-compl}

The goal of this section is to define a philosophical model of scientific model selection that is
based only on empirical accuracy and the complexity of the model's assumptions.  To define the
complexity of the assumptions of a scientific model I must first clarify what I mean by a
scientific model.  Requirements are kept to a minimum at this stage.

\begin{defn}
  \label{def:model}
  A {\bf model} is a tuple
  ${\cal M}=\{P,R,B\}$, where:
  \begin{itemize}
  \item $P$ is a set of assumptions\footnote{'Assumptions', 'Hypotheses', 'Principles',
  'Postulates' are used as synonyms in this paper.  Of course, there is no claim that they should
  be self-evident, consistently with the interpretation of, e.g., Jacobi, Mach, Riemann and many
  others \citep{pulte1998jacobi}.};
  \item $R$ is a set of results, which are logically derived from $P$;
  \item $B$ is a set of basic measurable quantities\footnote{Note that {\em quantities} can be
  more complex than natural language {\em terms} and they don't have model-independent
  meaning. There is in fact no conflict with the classic criticism of \citet{PutnamWTAN}. Note
  also that there is often a smooth transition between measurable and non-measurable quantities,
  whenever the expected precision of a quantity gradually deteriorates for some parameter ranges,
  until that quantity reaches a range where it is not measurable anymore.} that enter the $P$ and
    are assumed to be directly measurable with precision $\Delta(b)$ ($\forall b\in B$).
  \end{itemize}
\end{defn}

$P({\cal M})$ contains all the assumptions needed to deduce the results $R$ to be compared with
the experiments (including the rules of logic, all required mathematical assumptions, suitable
models of the experimental devices, approximations, background science, initial conditions,
tolerance $\Delta$ of all the quantities $B$ that we assume to be directly measurable\footnote{It
might be convenient to distinguish core assumption, that we rarely change, from auxiliary
assumptions that we often change (e.g. boundary conditions).  Correspondingly, classic models like
'Newton Gravity' can be seen as a family of models as defined here.}).  Any result in $R$ that
can't be derived from the assumptions must be part of the assumptions.

The above structure has some similarities to the much criticized received view \citep{Feigl}.
Hence, it is important to stress the key differences.  We can't assume observation sentences or
even properties that are theory independent.  The measurability of $B$, with a specific precision
$\Delta$, is part of the assumptions that can be tested only holistically \citep{Quine2Dogs,
  Quine2DogsRetro}.  No general and neutral observation basis is assumed and none is needed.
Still, theories can be tested against each other.  This is possible as long as a directly
measurable basis exists that {\em can be shared among those theories}.  Although
incommensurability \citep{sep-incommensurability} remains a theoretical possibility, there is no
evidence of two models dealing with the same topic where it is not possible to find a common set
of directly measurable quantities (see e.g.~\citet{fletcher2024alleged} for a recent study).

It is worth elaborating more on the differences between the present framework and the one of
\citet{CarnapAufbau}.  The idea of deriving basic observational properties from the observations
of similarities between elementary perceptions has proved impossible both because $n$-ary (for any
fixed $n$) similarity relations do not contain sufficient information \citep{leitgeb2007newQA},
and because we cannot {\em ensure} that anyone will see similar objects in the same way.  We can
share prototypical examples\footnote{E.g. we can show many examples on how to use a yardstick
under different circumstances.}, we can add narrative, but, no matter how much effort we put on
clarifying a concept, we can always only {\em assume} (never ensure) that any similarities will be
unambiguous.  All that we can do is to test those assumptions (but only partially and
holistically, together with the other model assumptions) and analyse statistically any unexpected
outcome. This is the motivation behind Def.~\ref{def:model}.

It is worth noting that I have not defined {\em direct measurement} explicitly.
Post.~\ref{def:P1}---together with Def.~\ref{def:model} below---offers an implicit (partial)
definition of direct measurements\footnote{One could define direct measurement explicitly as any
measurement that fulfills Post.~\ref{def:P1} and Def.~\ref{def:model}.  But this would designate
as 'direct' also measures that, intuitively, are not quite so. I find it clearer to leave to the
theory the responsibility of choosing which measurements are directs, but---whatever they
are---they must fulfill Post.~\ref{def:P1} and Def.~\ref{def:model}.}.  On the other hand, any
derived (i.e. non-direct) measurement must be constructible from a basis of direct ones.  It is up
to the model to decide which measurements are direct and which are derived, provided that
Post.~\ref{def:P1} and Def.~\ref{def:model} are fulfilled.  The exact separation between direct
and derived is largely conventional: in fact, I do not defend a unique objective cut between the
empirical data and the theoretical concepts.  But, it is important to ensure that every
measurements ultimately relies on something that fulfills Post.~\ref{def:P1}.

\paragraph{Equivalent formulations.}
As per Def.~\ref{def:model}, two different formulations of the same theory count as different
theories.  A core idea of the semantic view \citep{SuppesWST, FraassenImage} is that
representations of scientific theories should be independent on the language in which they are
formulated.  This idea is compelling, but the debate in \citep{glymour2013theoretical, van2014one,
  halvorson2013semantic} has made clear that, while some empirical constraints are necessary for a
sensible definition of theory equivalence, a precise proposal in that sense is not available,
within the semantic tradition.  The following is meant to fill this gap:

\begin{defn}
  \label{def:eq}
  ${\cal M}$ and ${\cal M}^{\prime}$ are {\bf equivalent formulations} (${\cal M} \equiv {\cal
    M}^{\prime}$) iff:
  \begin{enumerate}[label=\alph*)]
  \item there is some form of {\bf logical equivalence}\footnote{To be specific, I refer here to
  bi-interpretability \citep{Visser1991, visser2004categories, mceldowney2020morita} or,
  equivalently, Morita equivalence \citep{barrett2016morita, halvorson2019logic}.  However, the
  conclusions of this paper do not depend on the specific class of logical equivalence adopted.
  In fact, even a very restrictive notion of logical equivalence cannot exclude, by itself a form
  like $\Xi=0$.  On the other hand, even under very permissive equivalence, the following
  requirement (b) prevents the general identification of any theory with the $\Xi=0$ formulation.}
    between ${\cal M}$ and ${\cal M}^{\prime}$;
  \item there is a bijective mapping $J$ between the measurable properties of ${\cal M}$ and
    ${\cal M}^{\prime}$ such that for each measurable property\footnote{Measurable properties are
  the concepts of ${\cal M}$ that can be (operationally) defined from directly measurable
  properties.}  $p$ of ${\cal M}$, $J(p)$ is also measurable for ${\cal M}^{\prime}$ with the same
    precision and same outcome (via $J$). ({\bf empirical equivalence})
  \end{enumerate}
\end{defn}

It is well known that two models can be empirically equivalent while logically inequivalent
\citep{mormann1995incompatible}. For example, special relativistic mechanics is empirically
indistinguishable from classical Newton mechanics for phenomena whose velocities are much smaller
than the speed of light.  On the other hand, two models can be logically equivalent, but
empirically inequivalent.  This possibility is less discussed in the philosophical literature, but
it is quite obvious to the scientific practitioner.  For example, if I define the unit of length
based on my foot, rather than the modern reference in \citep{meterSI}, I obtain an alternative
model that is logically equivalent to the original one (the assumptions of the model are exactly
the same, except for what we chose to label as directly measurable), but significantly less
accurate than (hence empirically inequivalent to) the original one.  Note that the previous
discussion lets us conclude, in particular, that a $\Xi=0$ 'reformulation' of model ${\cal M}$ is
not, in general, empirically equivalent to the original model ${\cal M}$ and it is therefore {\em
  not} just a reformulation.

\paragraph{Epistemic complexity.}
The equivalence class of models that results from Def.~\ref{def:eq} is, finally, the object whose
complexity we must define, to make precise Einstein's intuition of the {\em complexity of the
  assumptions}. Indeed I can now define the epistemic complexity of a model ${\cal M}$ as the
minimum, over all equivalent formulations, of the length of its assumptions:
\begin{defn}
  \label{def:complexity}
  The {\bf epistemic complexity} ${\cal C}$ of a model ${\cal M}$ is the minimal length---across
  all possible equivalent formulations (in any language) of ${\cal M}$---of the assumptions
  $P({\cal M})$. In other words:
  \[
    {\cal C}({\cal M}) := \min_{{\cal M}^{\prime} \equiv {\cal M}} \mbox{length}\left[P({\cal
        M}^{\prime})\right].
  \]
  The {\bf epistemic simplicity or conciseness} of a model ${\cal M}$ is the inverse of its
  complexity.
\end{defn}

This definition is inspired to Kolmogorov-Chaitin (KC) complexity \citep{Kolmogorov,
  chaitin1975randomness, Zenil2020review}.  However---and this is the key difference---KC
complexity makes no reference to measurability.  So, it must be defined for a fixed, externally
given language.  Otherwise, there is always a language (or Turing machine) in which KC becomes
trivial (the $\Xi =0$ formulation discussed before).  The dependence on the language is fatal for
epistemological applications of KC complexity, because different choices allow any conclusion.

But, if I restrict it to {\em logically and empirically equivalent formulations}, I ensure that
$\Xi=0$ is not anymore a legitimate version of my original model and I have a definition that is
both {\bf non-trivial} and {\bf formulation independent} by construction (it only depends on the
choice of what I can measure, which is given by the nature of the problem and not by an arbitrary
choice).  This is not a small feat: it is {\bf the only proposal I know of a non-empirical
  candidate epistemic value which is precisely defined, non-trivial and as much
  formulation-independent as one can possibly wish}.

Note that epistemic complexity is {\bf defined precisely} but {\bf estimated approximatively}, as
it is the case for any empirical quantity in science).  As a result, it also explains scientific
disagreement as different estimates of the epistemic complexity of different models (see also the
discussion in the next paragraphs, as well as Sec.~\ref{sec:values} and
\citet{scorzato2026oxphos}).  This also explains why disagreements eventually settle, which is a
major puzzle for other accounts building on \citep{KuhnOVTC}.

Moreover, real scientific models are often too complex (having dependences with multiple domains)
to make a radical reformulation practical.  But, for the same reason, it is also doubtful whether a
radical reformulation will be able to achieve much greater simplicity.  That explains why the
scientists assess the epistemic complexity of a model by referring to the existing formulation in
common scientific language.  Hence, it {\bf justifies the use of ordinary scientific language to
  assess simplicity}.

\paragraph{Model selection.}
Now that we have at least one well defined non-empirical epistemic value which is non-trivial and
represents what we were looking for, can we build a philosophical model for scientific model
selection based on empirical accuracy and conciseness alone? This is detailed below.

How do scientists compare two scientific models, to decide whether any of them should be excluded?
If they are equally empirically accurate (i.e. both models describe---or fail to do so---all {\em
  existing} empirical evidence, with the same precision\footnote{Note that 'empirical equivalence'
defined in Def.~\ref{def:eq} refers to all conceivable {\em measurable} properties, while the
expression 'equally accurate' refers to the existing, already {\em measured} quantities.}), then
the scientists can directly compare the epistemic complexity of the two models and discard the
more complex model, if the difference is larger than the estimated uncertainties in the assessment
of the complexity (otherwise, both models are kept).

If the two models are not equally accurate, the scientists first identify the corrections (ad-hoc
assumptions) that would be needed to make the least accurate model as accurate as the best one.
If this requires too many or too complex corrections to the assumptions, it makes sense to drop
the most complex model, because it is just more complex for no empirical advantage.  For example,
imagine that models $M_1$ and $M_2$ are equally empirically accurate for most observations, except
that $M_1$ explains accurately experiment $e_1$ and $M_2$ doesn't, while $M_2$ explains accurately
experiment $e_2$ and $M_1$ doesn't.  The necessary ad-hoc assumptions in this case are $A_1$
($A_2$): ``the experiment $e_1$ ($e_2$) is not reliable''.  Hence, we can and must compare the
simplicity of $M_1 \bigwedge A_2$ vs $M_2 \bigwedge A_1$.  A single ad-hoc assumption is rarely
decisive (unless it is the only difference between $M_1$ and $M_2$, which is also often the case),
but the accumulation of evidence on one side, will make the cost of the multiple ad-hoc
assumptions unambiguously higher.  These difference will also motivate the scientists to devise
new experiments which cannot be conclusive by themselves, strictly speaking, but they become
conclusive if they succeed in forcing more and more complex ad-hoc assumptions predominantly on
one side.

\begin{figure}
  \centering
  \includegraphics[height=0.4\linewidth]{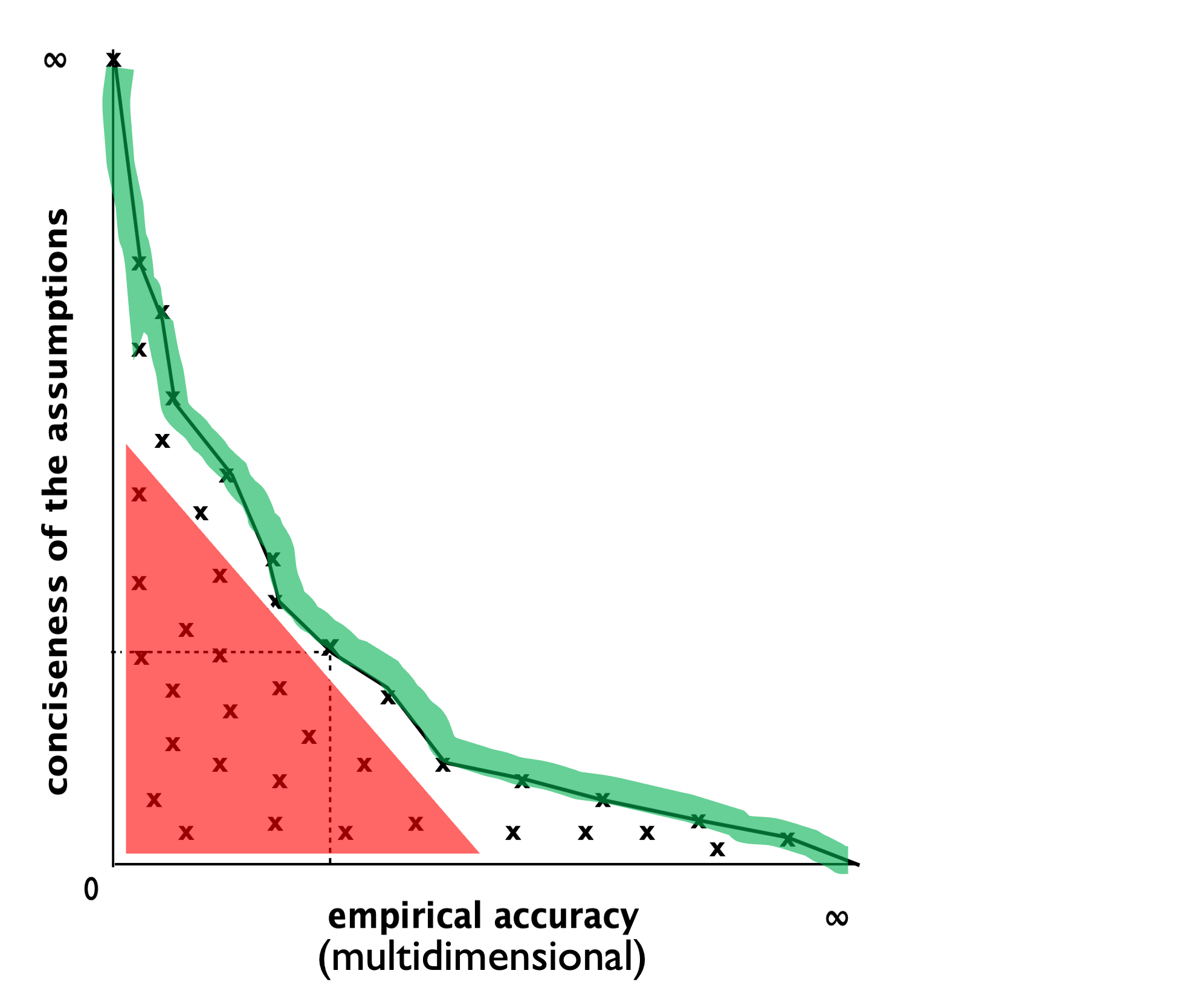}
  \caption{\label{fig:acc-con}Each 'x' represents a different scientific model, plotted by
    accuracy (which is multi-dimensional, but it is shown here as one-dimensional for simplicity)
    and conciseness. Ruled-out models appear in the red ares. State-of-the-art models are those in
    the green surface.}
\end{figure}

Note that the scientists typically make this selection unconsciously: when a model requires too
many corrections (ad-hoc hypotheses) and another model exists that needs no correction at all (or
only minimal ones), the former is not even considered an option: the selection is uncontroversial
and the scientists don't even realize that they are using epistemic complexity to exclude them.
But these models are legitimate from a logical point of view, they are infinitely more than
``good'' models, they entail different predictions and they are subtle to identify from the
philosophical point of view.  We can call them {\em ruled-out} models (the red area in
Fig.~\ref{fig:acc-con}) and defining a precise rule to exclude them is the focus of this work.

On the other hand, the models that are not worse than any other model represent the {\em
  state-of-the-art} (SotA) models (the green surface in Fig.~\ref{fig:acc-con}).  Different SotA
models represent different trade-offs between simplicity and accuracy in different applications.
Choosing among these models is often the focus of the scientists, and already well-defined
criteria exist to chose among them: the scientists may select only some models based on the
minimal precision required by a specific application and/or they may opt for a Bayesian model
average.  So, even if the State-of-the-art may be still very large, the selection between
different SotA models is not conceptually problematic and it is not the focus of the this work,
nor it is the concern of Goodman's New Riddle, because they are all legitimate (non-grue)
extrapolations and the remaining selection among them is scientifically well defined.
In summary, model selection, state-of-the-art models and ruled-out models are
defined as follow:

\begin{defn}
  \label{def:selection}
{\bf (Model selection)} A model $A$ is preferred to model $B$ if $A$ is neither more complex nor
less empirically accurate than $B$, while being strictly better (beyond uncertainty) than $B$ in
at least one of these aspects.  In this case, we say also that model $A$ is {\bf better} than
model $B$ and $B$ is {\bf worse} than $A$.
\end{defn}

\begin{defn}
  \label{def:sota}
The {\bf State-of-the-art} is the ensemble of models which are not worse than any other model.
The models that are not state-of-the-art are {\bf ruled-out}.
\end{defn}

Note that the {\bf scope} of a model is taken into account by multi-dimensional accuracy in the
following sense.  If the scope of the model $M$ is $D$ and $D'$ lies outside its scope, we can
conventionally assign very poor accuracy to $M$ on domain $D'$. Then, a model $M'$ that is less
accurate than $M$ on $D$, but covers also $D'$, is more accurate than $M$ on $D'$.  In this way,
we do not need to introduce scope as an independent epistemic value, because it is implicitly
included in a multi-dimensional notion of accuracy. This also reflects the idea that we do not
want to decide, at the philosophical level, whether we prefer an accurate model with narrow scope
or an approximate one with broader scope: they are both part of the state-of-the-art, and the
choice depends on specific requirements of particular applications.

\section{Why this solves Goodman's New Riddle and the problem of theory choice}
\label{sec:just}

What I have proposed is a philosophical model of scientific model selection, which solves
Goodman's New Riddle because I have argued that they are, essentially, the same problem.  In this
section I elaborate on the justification of this claim.  It is tempting to try to justify a model
of scientific model selection in terms of some superior principle, or in terms of likelihood of
future success. The temptation must be resisted, because superior principles will never be
superior for everyone, the likelihood will depend on further hidden assumptions and we would
simply end up with yet another inconclusive attempt to solve the {\em Old} Riddle of induction.

A {\em scientific} model of scientific model selection should be assessed and justified as any
scientific model, i.e., primarely on the basis of how accurately it describes its target
phenomena. Therefore, I {\em assume the epistemic value of conciseness as a working assumption}
and assess whether the concept of model selection based on it provides a good description of those
theory choices that enjoy a broad support by the scientific community.

A common misconception is that we already have descriptive accounts of scientific model selection.
Unfortunately this belief ignores the pervasiveness of the New Riddle discussed in
Sec.~\ref{sec:perva-new}.  I am not aware of any account that only relies on concepts that don't
become empty under deeper scrutiny, by exploiting the $\Xi=0$ trap.  I review some examples in
Sec.~\ref{sec:values}.  Only after solving the $\Xi=0$ trap it becomes meaningful to assess the
accuracy of a given account.

Multiple examples supporting the accuracy of the present model have been published already,
covering reductions and unifications \citep{Scorzato}, gravitational theories, quantum mechanics,
theories of evolution \citep{Scorzato-silfs14}, pseudo-science \citep{Scorzato-furbetto}, machine
learning models \citep{scorzato2024reliability} and the oxidative phosphorylation controversy
\citep{scorzato2026oxphos}.  A new example is described in Sec.~\ref{sec:conspiracy}.  To
challenge the philosophical model examined in this paper one should find at least one
counterexample, namely a model\footnote{The model should cover a {\em relevant} domain. The
present philosophical model does not try to determine which topics are relevant.}  that:
\begin{itemize}
\item either {\em should} be ruled-out and my philosophical model {\em doesn't},
\item or {\em shouldn't} be ruled-out and my philosophical model {\em does}.
\end{itemize}
Here, {\em should} and {\em shouldn't} refer to selections supported by broad scientific
consensus.  In other words, the proper test of this philosophical model must be performed with
cases of scientific model selection that are undisputed.  For those questions where there isn't a
clear scientific consensus, the present model offers an original prediction that the scientists
should probably consider \citep{Scorzato-furbetto}.

Before analysing a new example in Sec.~\ref{sec:conspiracy}, I answer two common questions. The
first is whether this model is descriptive or normative (Sec.~\ref{sec:possci}). The second is
about the choice of the specific notion of complexity introduced in Def.~\ref{def:complexity}
(Sec.~\ref{sec:simplicity}).

\subsection{Philosophy of science as a science itself}
\label{sec:possci}

A frequent question is whether my model is {\em merely} descriptive, given that it isn't justified
in terms of future likelihood.  I agree, if we agree that all our best scientific theories are
merely descriptive.  In fact, they do not enjoy any deeper justification beyond being concise and
descriptive.

It is important to observe that the present model is automatically {\em self-consistent}.  Namely,
it is itself a simple and accurate description of the domain that it tries to describe: scientific
model selection.  In fact, requiring conciseness is itself a concise rule. This is essential, to
recognize philosophy of science a science itself (science of science\footnote{In a sense that is
complementary to \citep{fortunato2018scisci}.}).  This property is also far from obvious.  For
example, saying that ``Scientific model selection is decided by the scientists'' is not
self-consistent, because the rule quoted here is fixed, and it does not leave room to the
philosophers to change it.  As another example, defining projectible models as those included in a
given list of models also cannot be self-consistent, without a vicious infinite regress.

Self-consistency alone is not sufficient, of course. For example, the rule ``Good theories start
with the word {\em Good}'' is also self-consistent, but not accurate.  However, if we accept that
the Old Riddle cannot be solved, then self-consistency, together with accuracy, is the best we can
aim for to justify the choice of the non-empirical values.  It is also a very strong requirement:
besides conciseness, I am not aware of any other non-empirical value able to achieve it.

\subsection{On the choice of the notion of epistemic complexity}
\label{sec:simplicity}

A common misconception is the idea that there are many possible definitions of simplicity and it
is therefore arbitrary to pick one.  If that were the case, we could select the definitions of
simplicity that describe best the actual (consensual) choices of the scientists: those
definitions would already provide better solutions to Goodman's New Riddle than anything else
proposed.  We do not have this option, because the real problem is not the {\em abundance} of
definitions, but rather the {\em absence} of any definition that survives the $\Xi$ trap.  Any
such definition is actually trivial under reformulation of the model and therefore totally
non-descriptive. Unfortunately, this point is rarely discussed in the literature, where
philosophers often criticize some proposals as {\em merely descriptive}, while they are not. In
fact, they are typically so vaguely defined that they might appear descriptive, but they are,
strictly speaking, just undetermined.

After solving the $\Xi$ trap, the next step is to adopt an epistemic value that captures the
hidden assumptions underlying actual scientific model selections supported by a broad scientific
consensus and accurately reproduces the scientists' conclusions.  In the rest of this paper, I
have defended the choice of Epistemic Complexity Def.~\ref{def:complexity} on these grounds. In
this section I consider potential alternatives.

For example, one may wonder why I did not chose other classic measures of complexity (e.g.
\citet{AIC, Schwarz78, sober2015ockham}).  What distinguish these measures (on one side) from
Kolmogorov-Chaitin and epistemic complexity (on the other side) is that the latter are very
generally applicable: they can be used to compare models that make totally different assumptions
(but describe the same phenomena). This is necessary to understand model selection across
revolutionary times, or to address potential challenges from pseudo-science, or simply to compare
models based on very different formulations. Measures like those of \citet{AIC} or
\citet{Schwarz78} are only defined within a given parametrization, because only there it makes
sense to compare the number of parameters.  On the other hand, if we restrict ourselves to compare
models based on the same parametrization, one can argue that epistemic complexity and \citet{AIC}
roughly agree, as long as the different parameters need to be introduced by different definitions,
assumptions or measurements.

Other notions of complexity do not capture the complexity of the assumptions and, therefore, they
do not capture what matters for the scientists to select a model.  For example, one could consider
the complexity of {\em deriving results} from a model (such as 'proof complexity'
\citep{krajicek2004proof} or the computational complexity to derive a prediction from a Neural
Network).  This category definitely matters to assess the opacity of a model
\citep{beisbart2021opacity}, but it does not introduce a new, independent, dimension valuable for
scientific model selection.  In fact, imagine that model ${\cal M}$ is as accurate as ${\cal
  M}^{\prime}$, ${\cal M}$ has simpler assumptions, while ${\cal M}^{\prime}$ enables simpler
derivation of results.  Would I ever select ${\cal M}^{\prime}$ over ${\cal M}$?  If the
advantages of ${\cal M}^{\prime}$ had enabled the derivation of more (accurate) results than
${\cal M}$, then ${\cal M}^{\prime}$ would be more accurate than ${\cal M}$, but since this is not
the case, by assumption, then the advantages of ${\cal M}^{\prime}$ are only hypothetical and
questionable.  In other words, the simplicity of the derivations is already taken into account by
the value of accuracy, for the extent that it is indeed a confirmed advantage. I don't know real
cases that contradict this conclusion.  Similar arguments can be made for other notions of
complexity.

It is possible that other notions of complexity exist that are different from the epistemic
complexity discussed here, but implies {\bf the same conclusions} of model selection, within
estimation errors. This would not challenge the present model: it would provide another
perspective on it, but it would be consistent with it.  On the other hand, a
different definition of complexity that leads to {\bf different conclusions} for model selection
is interesting only if one first identifies some cases where epistemic complexity leads to a model
selection that differs from the scientific consensus.  However, no such counter-example has been
published, until now, to the model already published in \citep{Scorzato}.  In fact, the claim
presented earlier in this Sec.~\ref{sec:just} remains unchallenged.

\subsection{History as a scientific discipline}
\label{sec:conspiracy}

I claim that the characterization of model selection discussed in this paper is very general.  A
common objection is that epistemic complexity makes sense for highly formal domains, like physics,
but less so in scientific domains where mathematics plays a less prominent role.  To answer those
criticisms, I consider, in this section, a field as removed as possible from highly mathematized
ones: historical science.  History is certainly a science and the task of the historian is to
formulate conjectures whose likely effects agree with the available documents.

A major challenge for philosophy of science, when applied to history, is clarifying what's wrong
with unwarranted conspiracy theories.  In fact, they are dismissed by the vast majority of
historians, but their empirical accuracy is usually not the problem: they are often designed to
agree with all the evidence and also to adapt quickly to any new evidence.  Moreover, saying that
conspiracy-theorists' assumptions are {\em unlikely} is also unsatisfactory.  Indeed, they
generally envisage circumstances that are very special by construction and, therefore, can't be
declared unlikely, because there are no statistical data either in favor or against them.

Consider the example of the {\em Bielefeld conspiracy}
\citep{wiki:Bielefeld_conspiracy}\footnote{A satirical theory plays here a useful role analogous
to a {\em thought experiment} in physics: it allows discussing the essential features of a
conspiracy theory without the interference of other complex factors that are inevitable in any
seriously meant conspiracy theory.}.  It claims that Bielefeld does not exist.  According to the
theory, if you say that you have never been in Bielefeld, you confirm that it might not exist. If,
instead, you say that you have been there, you must either be part of the conspiracy or have been
deceived by it, which also confirms its widespread penetration.  The theory always has an answer
to any counter-evidence.  The claim that it is {\em unlikely} that so many people lie or have been
deceived about having been in Bielefeld is not justified.  In fact, to explain its widespread
diffusion, the conspiracy calls upon alien forces and extraordinary hidden organizations which are
unique phenomena by construction.  No statistical evidence exists to either support or rule out the
claim.  Once again, accuracy and probability alone can't dismiss such conjectures. We can dismiss
them only because they require {\em extra and complex assumptions}, which are not necessary to
explain the evidence.  Hence, we face again the problem of quantifying the amount of assumptions.

Can I use the $\Xi$ trick to make a conspiracy theory as concise as the standard one?  It is very
instructive to see what happens in this case.  The proponents of the Bielefeld conspiracy
implicitly make the following two assumptions (or similar ones):

\begin{itemize}
\item In general, people lie with a probability of $<1\%$ [ordinary assumption].
\item Except, people who allegedly lived in Bielefeld, who lie all the time [specific ad-hoc
  assumption necessary to justify the conspiracy claim].
\end{itemize}

We can hide the complexity of the second assumption if we say that $\Xi$-people stands for anyone,
except those who allegedly lived in Bielefeld (who lie all the time).  Then the assumption of the
proponent of the conspiracy becomes as concise as the ordinary assumption:

\begin{itemize}
\item In general, $\Xi$-people lie with a probability of $<1\%$.
\end{itemize}

But to gain any advantage from this reformulation, one should then use {\em $\Xi$-people} rather
than {\em people} everywhere in history and the body of science.  But to preserve empirical
accuracy in this new formulation, one should ensure full convertibility between the concepts of
{\em people} and {\em $\Xi$-people} in every measurement.  In particular, any survey about any
topic should also ask whether the respondent lived in Bielefeld!  The same information should be
verified about any person who is part of any studies or plays any role in any topic.

Such measurements are not prohibited by any fundamental natural law---as opposed to those
involving chaotic systems discussed in \citet{Scorzato}---but they are nevertheless not available.
Proponents of the Bielefeld conspiracy cannot just claim that the notion of {\em $\Xi$-people} is,
logically, as legitimate as the notion of {\em people}: they should provide evidence of
measurements expressed in terms of {\em $\Xi$-people}.  Just as a set of emeralds, whose first
discovery was not recorded, is not convex under the grue property (see Sec.~\ref{sec:direct-meas}
and Footnote \ref{fn:convex}), a set of people whose stay in Bielefeld was not recorded is also
not convex under the property of $\Xi$-people.

In conclusion, even for domains that rely on minimal mathematical background, gaining conciseness
artificially is logically possible, but only by compromising accuracy, which means that the
conciser model is not empirically equivalent to the original one.

\section{Comparison with previously proposed solutions and their shortcomings}
\label{sec:discuss}

\subsection{Relation to previously proposed solutions}
\label{sec:prev-sol}

Since Goodman posed his riddle 70 years ago, many solutions have been proposed.  The goal of this
section is not to review comprehensively the huge literature on this topic, but rather to compare
the present model to those proposals that have interesting similarities and differences.

The idea that grue-like concepts are {\bf not observable} has been put forward very early
\citep{grue83}, but it is not correct: to observe the grueness we simply need a detector of colors
and a clock.  Here, consistently with \citet{Gardenfors_1990, Gardenfors_2000}, I have emphasized
what distinguishes {\em direct} observations from general observations.

Fodor's idea \citep{Piattelli1980} that some hypotheses are {\bf innate} was already dismissed by
Putnam as a potential solution \citep{grue83}.  In particular, scientists often introduce new
hypotheses and concepts that cannot be considered innate, but they are nevertheless very
successful (e.g. quarks).  On the other hand, feasible and accurate {\em measurements} have the
right degree of flexibility: they are neither fixed nor arbitrary. Scientists are able to design
new experimental devices, but doing so is not as easy as introducing a new grue-like concept.  The
constraint of what is measurable is also a natural one for scientific models.

A vast literature \citep{NatKind, NatKind-IEP} has tried to characterize the notion of {\bf
  natural kinds} and the related notion of {\bf similarity} \citep{QuineOntoRel}.  Identifying the
right concept of similarity is fraught with issues \citep{fletcher2016similarity}.  The original
idea was that only natural kinds are used for projectible laws. The program attracted intense
research over half a century, but it proved too ambitious. I will not review the multiple dead
ends the program ran into, as it is done very well by \citet{NatKind} and \citet{NatKind-IEP}.  I
will just explain very briefly why I believe that the project itself is not a good idea.  On one
hand, it seems exceedingly difficult to be able to tell why the concept of {\em quark} (which lies
at the heart of our best scientific theories) should be more natural than many grue-like
concepts\footnote{Except if we {\em use} the fact that the concept of quark does appear in our
best scientific theories. But then we can't use the concept of natural kind to tell what {\em
  should} be projectible, which misses the point of why natural kinds were introduced in the first
place.}.  On the other hand, it is still possible to build accurate, but completely implausible
models based only on very natural kinds. This is easily accomplished by formulating very crude
general laws, accompanied by a long list of exceptions.  In this respect, \citet{Gardenfors_1990}
introduces a key refinement of the definition of natural properties, but he doesn't go beyond the
idea that essentially identifies what is natural with what is projectible. This identification is
problematic, because it leads to a criterion that is both too strong and too weak as explained
above and in the end of Sec.~\ref{sec:direct-meas}.

My proposal can be seen as limiting the role of natural kinds to where it is strictly necessary:
only for properties that we consider directly measurable. The criterion for admissibility is
consistent with the one proposed by \citet{Gardenfors_1990}: natural properties are convex, at
least locally\footnote{Requiring convexity only in a small region around a measurement also
answers the criticism that natural (or measurable) properties might not be globally convex
\citep{HernandezC2017}, although the specific example provided there is not correct
\citep{gardenfors2019reply_conde}. Note that the difference between a generic convex set and an
error-box is inessential, for small regions.}, but natural (or directly measurable) properties are
not enough to characterize what is projectible.  The other essential component is conciseness.
But directly measurable properties effectively introduce constraints on the language that enable a
non-trivial definition of conciseness.

Goodman's own theory of {\bf entrenchment} has been criticized by many authors
\citep{Stalker1994grue, Elgin1997goodman, sep-goodman}.  Some of them \citep{sep-goodman,
  Scholz2024CC} consider his solution merely descriptive.  This is not true.  If it were so, it
would be exactly what Goodman was looking for.  But it is not, as many others have pointed out
\citep{Teller1969projections}.  It is important to review why it is not.

First, it is very unclear how entrenchment is supposed to be assessed: (i) when does a past
hypothesis count as the same hypothesis? The same sentence is not the same hypothesis, strictly
speaking, when combined with different other assumptions, consistently with the idea of holism.
If we adopt this strict view, we preclude any useful application of entrenchment.  If we don't, we
must define a similarity measure among different hypotheses, which was not addressed.  Even if we
succeed, (ii) how do we count how much an old hypothesis was used successfully? Do my home
experiments (that I can repeat thousands of times per day) count as much as an experiment
conducted at CERN after twenty years of preparation?  This is a reformulation (not a solution) of
the problem of confirmation \citep{sep-confirmation}.  Secondly, even if all these major
uncertainties were settled, the model would still be wrong even in those cases where its
interpretation is rather unambiguous.  In fact, sticking with an old model and adding many ad-hoc
corrections to it would be clearly more entrenched than introducing a completely new simple and
accurate model.

There is, however, also some truth in Goodman's theory: past successful models do carry a legacy,
but not in the sense that they should be preferred, ceteris paribus, to more recent ones.  The
legacy exists because past successful models determine which features we decide to record, and
because old measurements remain a reference to asses any new model, as discussed in
Sec.~\ref{sec:direct-meas} and Sec.~\ref{sec:conspiracy}.  Although this might (or might not) be
an advantage for the older model, it does not introduce arbitrariness into the comparison, because
it is a fact that is hard to change.

One of the most popular modern approaches to confirmation theory \citep{sep-confirmation} is {\bf
  Bayesianism} \citep{Bayesian, SprengerHartmann-bayesian}.  However, the outcome of a Bayesian
analysis depends on the choice of the prior probabilities (aka priors), which remain relevant for
any realistic amount of data and any non-toy application \citep{sep-science-theory-observation,
  norton2021material, scorzato2024reliability}.  In turns, the priors can only be defined by
relying on some non-empirical criteria, which is vulnerable to the $\Xi$ trick, unless we adopt a
non-trivial, reformulation independent measure of complexity, whose only published option is
Def.~\ref{def:complexity}.  If we do that, it makes no sense to keep ruled-out models (according
to Def.~\ref{def:selection}) in the Bayesian mix, because no evidence can ever prefer them over
some state-of-the-art model, but they add unbounded perturbations to the Bayesian outcome.  Hence,
also Bayesianism can make sense only if its definition relies on epistemic complexity, which then
makes it equivalent to the model defended here.

The very influential {\bf material theory of induction} \citep{norton2021material} accepts that
all inductions must rely on additional specific assumptions that depend on the domain under
consideration (i.e. they are {\em local}).  This aligns with the present emphasis that we are
typically interested in extrapolating entire scientific models rather than individual sentences.
But the analysis fails to identify those assumptions that the scientists do not make explicit, but
are essential to select meaningful laws.

To avoid confusion, it is also worth mentioning that a significant branch of philosophical
literature (see e.g. \citet{okasha2007jackson} and references therein) interprets Goodman's New
Riddle as an internal {\bf contradiction strictly within the syntactic/formal confirmation theory}
of \citet{hempel1943purely}.  A 'solution', in that interpretation, is one that eliminates the
contradiction. I have offered extensive evidence in Sec.~\ref{sec:perva-new} that Goodman's New
Riddle still represents a thoroughly contemporary and general challenge for any modern approach in
the philosophy of science.  Specfically, the riddle is not so much about removing a contradiction;
it is rather about removing a symmetry between plausible and implausible models.

\vskip 0.2 cm

Because I have emphasised that the problem of {\bf theory choice} is the proper generalization of
the New Riddle of induction, any attempt to solve the former is also relevant to solve the latter.
However, in recent decades, the philosophical literature on theory choice has largely swept the
New Riddle under the rug.  For example, a very influential tradition, going back at least to
\citet{KuhnOVTC} (see \citet{de2024theory} and references therein for recent developments), argues
that theory choice is determined by a bunch of vaguely defined and evolving values.  However,
these values are never fully defined.  In particular, non-purely empirical values like simplicity,
parsimony, external coherence and explanatory power always appear in such lists, but in absence of
a precise definition, it is not clear how they could possibly avoid the $\Xi$ trap.  Hence, all
those accounts remain fundamentally undetermined and therefore non-descriptive.  This topic is
discussed further in Sec.~\ref{sec:values}.

A further instructive example, that lies outside the main stream, but still suffers from similar
shortcomings, is the proposal by \citet{Dawid2015-DAWTNA}.  In this case, a theory is selected if
it has 'no alternatives'.  However, this idea makes sense only if we assume some non-empirical
constraints, otherwise we always have infinite alternatives (see Sec.~\ref{sec:ts-prob}).  This
fact is indeed recognized be the authors \citep{Dawid2015-DAWTNA}.  But, without a clear
definition of the constraints, also this model remains vulnerable to the $\Xi$ trap and
undetermined.

Finally, the goals of \citet{douven2020natural} are very different from the present ones: they aim
at a model for natural concepts in a socio-cognitive context.  However, they also rely on
convexity and parsimony.  Parsimony is not defined precisely there, and the realisation that
convexity enables a precise and robust definition of parsimony would certainly benefit also that
approach.

\subsection{Conceptual spaces}
\label{sec:con-spa}

I have already emphasized the deep relation between the present proposal and the one of
\citep{Gardenfors_1990, Gardenfors_2000}.  In this section, I elaborate more on this relation,
focusing on the interesting analysis presented in \citet{GardenforsStephens2017}.  The authors
distinguish three types of knowledge: {\em knowledge-how}, {\em knowledge-that} and {\em
  knowledge-what}, that correspond, respectively, to three different types of memory: {\em
  procedural}, {\em semantic} and {\em episodic}.

I acknowledge that the distinction plays an important role in understanding human cognition.
However, science differs significantly from natural human cognition: it may share the same basic
functionalities, but it is far from instinctive, it involves additional deep conceptual
elaboration and it is often unnatural for humans.  Importantly, science strives to reduce all
knowledge to knowledge-that, and it is mostly successful in this: the documentation of an
experimental setup is an excellent example of removing any dependencies on any informal know-how
and ensuring that the process is fully reproducible.  Other authors \citep{williamson2001knowing}
have argued that knowledge-how can be fully reduced to knowledge-that, and I agree with their
conclusions.

Matters are different, however, when it comes to {\em knowledge-what}.  While scientists still
strives to reduce, as much as possible, any knowledge-what to knowledge-that, there are
fundamental limitations that preclude a full reduction.  For example, the correlation between the
dimensions that characterize an apple can be expressed in terms of formal propositions that state
statistical correlations.  Moreover, whenever precision matters, the scientists will not rely on
the intuitive notion of apple, but rather on genetic analysis.  However, {\em some} basic
measurements cannot be expressed in terms of propositions that can be verified as true or false
(i.e. knowledge-that), except by introducing vicious circles.  The determination of color can be
reduced to a measurement of light wavelengths, but then we must assume a model for the
spectrometer and what we measure directly is simply shifted from the relation eye-apple to the
relation eye-spectrometer display (and many other direct measurements for set-up and calibration).
In other words, there is an irreducible core of knowledge-what that is represented by the
fundamental direct measurements that any model must assume somewhere.  Science tries to reduce
their scope to quantities that are as unambiguous as possible, but even if we could reduce every
direct measurement to the reading of a digital display that prints either {\rm 0} or {\rm 1} on
the screen, we should still assume that different people at different times will interpret those
slightly different symbols in very different contexts in the same way.  This is where the dream of
a provably unambiguous observational language fails and must be replaced with assumptions about
the appropriate knowledge-what and its learning mechanisms, which can be verified only
holistically. Pure logic alone cannot fix the intended meaning of our terms: we can only emphasise
where we need to introduce an assumption.

In other words, the present view is consistent with \citep{GardenforsStephens2017}, but with the
caveat that the scope of knowledge-what, in science, is reduced to the essential.  This reduces
the need of convexity to small neighborhood around measurable points.  This point of view also
mitigates the implications of the criticism brought forward by \citet{Strossner2022} to the
applicability of conceptual spaces to natural multi-domain concepts.

\subsection{Do we need other independent epistemic values, besides accuracy and conciseness?}
\label{sec:values}

In this section I compare the present proposal to \citep{douglas2013value}, which represents a
significant step towards rationalization, with respect to the vagueness of \citep{KuhnOVTC}.  I
fully agree with \citep{douglas2013value} that {\bf internal coherence} and {\bf empirical
  accuracy} (concerning the past) are not negotiable.  Concerning internal coherence, the matter
is not undisputed.  So, it is worth reminding that {\em ex falso quodlibet}, which includes any
statement in conflict with the evidence.  Internal coherence must and can be recovered\footnote{In
this paper, I don't emphasise internal coherence, but it is assumed in
Def.~\ref{def:model}.}. This is typically done by adopting more complex assumptions. For example,
if $X$ and $Y$ are inconsistent, in general, we may say that in context $A$ we assume $X$, in
context $B$ we assume $Y$. So, potential contradictions are eliminated at the cost of higher
epistemic complexity.

While there is little doubt that these values are the top priority for science (let's call them
the {\em first class}, to emphasise the imperfect analogy to Douglas' {\em first group}), they are
far from sufficient to account for the scientists' consensus on theory choice, as discussed in
Sec.~\ref{sec:ts-prob}.  The second class of values recognized in \citep{douglas2013value}
includes: simplicity, scope, explanatory power, fruitfulness, unification, external coherence,
novel predictions and precision.  Note that the values of {\bf scope} and {\bf precision} are
included in my (multi-dimensional) definition of empirical accuracy.  So, I see them in the first
class.  In fact, I do not want to choose between a model that is more precise and a model with a
broader scope: I keep both in the state-of-the-art.

Before analysing the remaining values (the {\em second class}), it is worth discussing how we
expect them to be used.  One possibility is to imagine that the scientists unconsciously keep
weights in their heads, which they use to prefer one model over another and to quarrel with their
colleagues.  In this scenario, the discussions and further evidences lead to adjustments of the
weights until they converge.  This scenario does not explain why this social dynamics eventually
converges and does not move much afterwards.  Nor it explains why the scientists don't usually
recognize any subjective element in these personal weights and often stubbornly insist that their
conclusions are 'objective'.  But, this seems the dominant view among modern philosophers of
science.

Another possibility---defended here---is that, while there is certainly tension between the first
and the second class of values, nevertheless (a) there isn't much tension among the values within
the second class, when properly understood: they pull on the same direction; and (b) once we
recognize the single second-class dimension that matters, theory selection becomes rather
unambiguous, because the second class is used to pick the unambiguous best from what is left
undetermined by the first class.  Both (a) and (b) are essentially defended also in
\citep{douglas2013value}, but both claims remain necessarily vague there, because all the values
in the second class (except novel predictions; more on this below) are vulnerable to the $\Xi$
trap and it is impossible to recognize how they are all proxies of a well defined underlying
value.  But, if we accept the solution to the $\Xi$ trap discussed in this paper, they can all be
defined and, as suggested in \citep{douglas2013value}, they do not pull in many different
directions.

After this premise on the role of the second class,
let us consider in more detail why the second class does not bring any value that is necessary and
truly {\em independent} from accuracy and conciseness.  First consider {\bf fruitfulness}.  If a
model has {\em proved} to be more fruitful, it must have already achieved some other advantage,
either in accuracy or something else, that must be specified and would refer to some other value
discussed here. Otherwise, its fruitfulness is only a conjecture that does not have to be included
in the current assessment of the model.

Let us consider {\bf external coherence}\footnote{The value of {\bf fundamental justifications} is
not discussed in \citep{douglas2013value}, because, I believe, it is a special case of external
coherence.}. Logical coherence between {\em different} models in the state-of-the-art is not
required and it is often violated: different models in the state-of-the-art are often competitors.
We can define external coherence as follows. Two model are largely externally coherent, if they
have a large common core of assumptions.  Then, the model of scientific model selection presented
here favors models that have larger external coherence with other state-of-the-art models, as it
is illustrated by the following example. Imagine that we need a model that describes precisely the
class of phenomena $E_1$ (e.g. rockets), but it should also be approximatively consistent with the
broad class of phenomena $E_0$ (e.g. general phenomena in physics and chemistry).  Imagine that
model $M_0$ is generally used to describe the broad class $E_0$, while the models $M_1$ and $N_1$
are two competing alternatives to describe $E_1$.  Imagine that $M_1$ is obtained from $M_0$ by
adding just one simple assumption, while $N_1$ is very different from either $M_0$ or $M_1$.
Although $M_1$ is more complex than $N_1$, because it includes $M_0$, $N_1$ alone is not
sufficient to describe $E_0$, and we can use it for our purposes only if we also assume $M_0$
anyway, to describe $E_0$.  So, $M_1$ is simpler than the combination of $N_1$ and $M_0$, because
it is more externally coherent with $M_0$ and it is therefore preferred.  Hence, external
coherence does not add any valuable {\em independent} dimension that is not already included in
the values of the first class plus conciseness.

Furthermore, {\bf unifications} necessarily represent either an increase in the scope of a model
or a reduction of its overall assumptions (or both).  Finally, {\bf explanatory power} would
require a much longer discussion, given the variety of meanings that have been associated to it
\citep{sep-scientific-explanation-20th}, but at least one of its most influential interpretations
\citep{Kitcher1989} is fully aligned with the idea of unification and with the reduction to
concise assumptions.

In conclusion, I do not see any evidence of the need to include other values in a model of theory
choice that are not already sufficiently taken into account by the values of the first class plus
conciseness\footnote{One more note about {\bf novel predictions}.  The model described in this
paper does not attribute extra value to the fact that a theoretical prediction was formulated {\em
  before} the first measurement of the predicted phenomenon.  Novel predictions have certainly
played a role, historically, to accelerate the adoption of a new theory. But, are they necessary
to define theory selection? Defining exactly {\em when} a phenomenon was first predicted and when
the {\em same} phenomenon was first observed, is fraught with subtleties.  The question is, do we
need to enter these subtleties? To argue that we must, one should provide an example where
accuracy and conciseness alone are not sufficient to justify a consensual case of theory
selection, but the time order of the respective proposals and observation achieves that goal.  I
am not aware of any such case and I doubt there are, because such theory selection would be very
fragile: the conclusion ought to change in case one would discover that some scientist was aware
or was not aware of early experiments.  Novel predictions are certainly impressive in the
transition phase that leads to new theory selections, but when a new theory has firmly established
its new role, it rarely depends on the time ordering of these events.}.  Moreover, this model
explains well why the scientific controversies mostly reach consensus: the debates are due to
different, legitimate, estimates of the complexity of hypotheses.  The complexity gap becomes
increasingly clear under the growing constraints arising from the accumulation of empirical
evidence.  Complexity of the assumptions and empirical evidence are also the topics that the
scientists actually discuss.

\section{Conclusions}

The solution to Goodman's New Riddle of induction discussed in this paper is based on a
combination of two main insights: (i) conceptual spaces \citep{Gardenfors_1990}---applied only to
directly measured concepts; (ii) the theory of complexity \citep{Scorzato, chaitin1975randomness,
  Kolmogorov}, which is used to characterize---in a formulation-independent way---the complexity
of the assumptions of a model.  In one sentence, {\bf the constraint of convexity enables a
  non-trivial notion of complexity}.

This provides a well defined model that makes it precise the informal idea that science always
aims at {\em explaining more with less}.  In the spirit of Goodman's {\em New} Riddle, I do not
try to explain {\em why science is successful}, I rather clarify {\em what we mean by science},
i.e., I identify the hidden assumptions behind the scientists' decisions about scientific model
selection.  I do not pretend to explain why those decisions must be right.

Most philosophers of science, today, believe that Goodman's New Riddle cannot be solved and many
also believe that it is a false problem.  Concerning the latter point of view, I have shown that
many of the most important open problems in philosophy of science lead to it, consistently to the
original Goodman's intuition, that is still valid.  On the other hand, the belief that Goodman's
Riddle cannot be solved, is implausible for multiple reasons.  First, it amounts to saying that
the conclusions of science are based on fundamentally ineffable assumptions\footnote{This is far
worse than admitting that all conclusions of science are based on assumptions that we cannot
conclusively test, which is true and recognized by everyone, except for the philosophically most
naive.}. This claim implies the futility of trying to clarify the exact assumptions behind any
scientific conclusion, which is one of the main focus of scientists of all disciplines.  The only
way to make sense of science is to admit that the scientists do use some hidden assumption, but
they are identifiable. Another reason why the skepticism is implausible is that some natural
options had never been explored seriously before.  In particular, although both {\em measurements}
and {\em complexity} are classic topics in philosophy of science, I am not aware of any work
(except for \citet{Scorzato}) that tries to define them in combination.  It is very natural to
expect that complexity makes sense in science only after introducing a constraint of what is
measurable with some stated precision.  How can we claim that Goodman's New Riddle cannot be
solved before exploring such a natural approach in full depth?

My philosophical model won't be the last word on the goals of science, but it is certainly the
best description currently available: it achieves excellent accuracy (I am not aware of any
counterexample) at the cost of a moderately higher level of sophistication than usual.  To
criticize it, or improve it, one should first identify at least one counterexample, namely a model
that:
\begin{itemize}
\item either {\em should} be ruled-out (according to scientific consensus) and my philosophical
  model {\em doesn't},
\item or {\em shouldn't} be ruled-out (according to scientific consensus) and my philosophical
  model {\em does}.
\end{itemize}

In the age of AI, the role of philosophy of science is more crucial than ever to assess the
reliability and to guide the evolution of models that represent a radical break with the tradition
of scientific modeling \citep{scorzato2024reliability}.  But this requires a philosophy of science
that adopts for itself the same standards that it identifies for all scientific disciplines.

\paragraph{AI disclosure.}
For the present paper Gemini AI was used to improve the clarity of the language, to search the
literature and to check the correctness of the main argument.

\appendix
\section{Legitimacy of the \texorpdfstring{$\Xi=0$}{trivial} formulation}
\label{sec:app_xi}

A few details are appropriate to explain why I claim that we can always reformulate the
assumptions of any model as $\Xi=0$, why this formulation is logically equivalent to the original
formulation and why it is legitimate to talk about measurements of $\Xi$ and their associated
error-boxes. See also \citet{Scorzato}.

The assumptions of any model are, in general, statements (assumed to be true), equalities or
inequalities. Any (in)equality can also be expressed as a statement whose value can be {\tt true}
or {\tt false}. In fact, the equation $A=B$ (or $A<B$) can be replace by a sentence $\Sigma$ that
is true if and only if, indeed $A=B$ (or $A<B$). On the other hand, we can also expressed any
sentence $\Sigma'$ as an equality, simply by defining $A$, such that $A=0$ if $S'$ is true and
$A\neq 0$ if $\Sigma'$ is false.  Moreover, any set of equations can be brought to the form
$\Xi=0$, by bringing any non-zero term to the left and assigning the symbol $\Xi$ to represent
everything on the left.  It might be challenging to interpret $\Sigma$ (or $\Xi$), but we have no
obvious clear-cut arguments to refute a supporter of the model who claims to be able to interpret
$\Sigma$ (resp.~$\Xi$) directly.  Developing that argument is the topic of this paper.

The claim that the new formulation is logically equivalent to the original one is also easy to
justify. Above, I have shown how to translate the original formulation into the new one.  To go
back to the original formulation it is sufficient to introduce, in the new formulation, as many
symbols as in the original formulation, identify them with the corresponding symbols in the
original formulation and assume the relation between $\Sigma$ (resp.~$\Xi$) and the newly
introduced symbols to reproduce the relations that we defined in the original formulation.  In
this way, the two formulations can be simply identified one-to-one.  Note that the introduction of
all these symbols in the new formulation does not increase the complexity of the new formulation,
because they are only needed to translate into the legacy formulation and they are not necessary
to formulate all the assumptions of the model in its most concise form.

If we claim that ``$\Sigma$ is {\tt true}'' or ``$\Xi=0$'' represent all the assumptions of the
model, we must also explain how are the measurements documented in this formulation.  This is
where things become challenging for a defender of the new formulation of the model.  But let's see
how far we can go.  Because the only symbol available is $\Sigma$ (resp.~$\Xi$), measurements must
be expressed in terms of that symbol alone.  Real measurements contain more information than just
a {\tt true} or {\tt false} outcome.  So, simply assigning a truth value to $\Sigma$ is not
sufficient: we should at least assign a probability to each possible outcome.  This leads us to
introduce at least another variable $P(\Sigma)$ that represents the probability of the outcome and
takes continuous values.  Because $P$ is continuous, we can discuss convexity.  Crucially, we
cannot prove that any measurement in the original representation is mapped into a convex interval
in the new representation $P$.  However, this is not the notation chosen in this paper.

In the $\Xi$ formulation we can assume that $\Xi$ itself can take continuous values. If $\Xi$
represents a sentence, we can, e.g., assume that $\Xi=0$ stands for certainly {\tt true}, $\Xi=1$
stands for certainly {\tt false} and the values in between stand for intermediate estimates of the
corresponding probability.  If $\Xi$ represents a set of equations, the value of $\Xi$ represents
how precisely the measurement has verified the relation. Using standard scientific notation, we
can write $\Xi = \Xi_0 \pm \delta$ to represent the statement that a measurement has found $\Xi$
to be within an interval of size $2\delta$ around $\Xi_0$ with a given probability.  Again, we
cannot prove that convex intervals in $\Xi$ represent convex intervals in the original
formulation.  In the main text, I refer to the formulation $\Xi=0$ for simplicity, because I find
it more intuitive and it does not require the introduction of new symbols.

One might also ask whether $\Xi=0$ represents a single equation or a set of equations. Both
options are possible, but both options have difficulties in representing measurements faithfully
and concisely.  In fact, I could certainly write the assumptions as $\Xi_n=0$, for $n=1,...N$,
which is only slightly more complex than just $\Xi=0$.  But, if I want to express the expected
error-boxes in a way that is faithful to the actual measurements in the original formulation, I
will be forced to write $\Xi_n \in \cup_j \Delta_{j,n}(\Xi_1, ...\Xi_N)$ with very complex
set-valued functions $\Delta_{j,n}$ that won't be simpler than the original formulation.

\section{Convexity and discrete measurements}
\label{sec:app_discrete}

It is not difficult to extend Postulates \ref{def:P1} and \ref{def:P2} to the case when the set of
possible measurement outcomes $Q$ is discrete.  This is necessary, for example, to describe
measurements whose possible outcomes are {\tt true/false}, integer numbers or other finite set of
categories.

A simple way to extend Post.~\ref{def:P1} and \ref{def:P2} consists in choosing a metric $d(.,.)$
in the space of $Q$ and define an error-box around a given value $Q_0$ as the set of all $Q$ such
that $d(Q_0,Q) < \Delta$.  The level set $\{Q:P(Q)\geq l \}$ and the concept of convexity are
still well defined on a metric space \citep{KhamsiKirk2001MetricSpaces} and Post.~\ref{def:P2} is
still meaningful and can be extended without change.

It is important to note that the choice of $d()$ is not arbitrary: $d(Q_1,Q_2)$ must represent how
unlikely it is that a measurement device could read $Q_1$ while the target system is in the state
$Q_2$.  Both over-estimating and under-estimating $d()$ negatively impacts the accuracy of the
model (either because the model claims lower precision than it actually has or because any natural
fluctuation appears as significant model failure).  The metric $d()$ forces us to embed the
discrete measurement outcomes into a continuum space that represents more faithfully the
underlying phenomena.  For example, the integer digits on the display of the experimental device
are typically discretizations of a continuum underlying process.  In this case, the digit 7 must
be assumed to be closest to digits 6 and 8.  However, if the experimental set up includes a steps
where the digits are handwritten, then we must also consider the possibility that digit 7 might be
close to the digit 1. In this case, the relevant underlying continuum space is the one of all the
possible handwritten digits.

In conclusion, discrete sets do not undermine the relevance of convexity in measurements, because
even if the outcome is discrete, we must identify the continuous range of possibilities that might
generate different outcomes in order to assign error-boxes to discrete measurements.  This is how
convex sets still play a fundamental role.

\bibliography{philo, ML}{} \bibliographystyle{chicago}

\begin{thebibliography}{}

\bibitem[\protect\citeauthoryear{Akaike}{Akaike}{1973}]{AIC}
Akaike, H. (1973).
\newblock {Information Theory as an Extension of the Maximum Likelihood
  Principle}.
\newblock In B.~Petrov and F.~Csaki (Eds.), {\em Second International Symposium
  on Information Theory}, pp.\  267--281. Budapest: Akademiai Kiado.

\bibitem[\protect\citeauthoryear{Baker}{Baker}{2022}]{Baker-SEP}
Baker, A. (2022).
\newblock {Simplicity}.
\newblock In E.~N. Zalta (Ed.), {\em Stanford Encyclopedia of Philosophy\/}
  (Summer 2022 Edition ed.). Stanford University.

\bibitem[\protect\citeauthoryear{Barnett}{Barnett}{1950}]{EinsteinLife}
Barnett, L. (1950, Jan).
\newblock {The Meaning of Einstein's New Theory -- Interview of A. Einstein}.
\newblock {\em Life Magazine\/}~{\em 28}, 22.

\bibitem[\protect\citeauthoryear{Barrett and Halvorson}{Barrett and
  Halvorson}{2016}]{barrett2016morita}
Barrett, T.~W. and H.~Halvorson (2016).
\newblock Morita equivalence.
\newblock {\em The Review of Symbolic Logic\/}~{\em 9\/}(3), 556--582.

\bibitem[\protect\citeauthoryear{Behroozi}{Behroozi}{2022}]{behroozi2022rectangleinconvex}
Behroozi, M. (2022).
\newblock {Largest Inscribed Rectangles in Geometric Convex Sets}.

\bibitem[\protect\citeauthoryear{Beisbart}{Beisbart}{2021}]{beisbart2021opacity}
Beisbart, C. (2021).
\newblock {Opacity thought through: on the intransparency of computer
  simulations}.
\newblock {\em Synthese\/}~{\em 199\/}(3-4), 11643--11666.

\bibitem[\protect\citeauthoryear{Bird and Tobin}{Bird and
  Tobin}{2008}]{NatKind}
Bird, A. and E.~Tobin (2008).
\newblock {Natural Kinds}.
\newblock In E.~N. Zalta (Ed.), {\em Stanford Encyclopedia of Philosophy\/}
  (Spring 2024 Edition ed.). Stanford University.

\bibitem[\protect\citeauthoryear{Boyd and Bogen}{Boyd and
  Bogen}{2025}]{sep-science-theory-observation}
Boyd, N.~M. and J.~Bogen (2025).
\newblock {Theory and Observation in Science}.
\newblock In E.~N. Zalta and U.~Nodelman (Eds.), {\em The {Stanford}
  Encyclopedia of Philosophy\/} ({S}pring 2025 ed.). Metaphysics Research Lab,
  Stanford University.

\bibitem[\protect\citeauthoryear{Brzovi\'{c}}{Brzovi\'{c}}{2014}]{NatKind-IEP}
Brzovi\'{c}, Z. (2014).
\newblock {Natural Kinds}.
\newblock {\em The Internet Encyclopedia of Philosophy\/}~{\em ISSN-2161-0002},
  1.

\bibitem[\protect\citeauthoryear{Carnap}{Carnap}{1950}]{Carnap-conf}
Carnap, R. (1950).
\newblock {\em {The Logical Foundations of Probability}}.
\newblock Chicago: University of Chicago Press.

\bibitem[\protect\citeauthoryear{Carnap}{Carnap}{1966}]{CarnapAufbau}
Carnap, R. (1966).
\newblock {\em {Der Logische Aufbau der Welt}\/} (3rd ed.).
\newblock Hamburg, Germany: Felix Meiner.

\bibitem[\protect\citeauthoryear{Chaitin}{Chaitin}{1975}]{chaitin1975randomness}
Chaitin, G.~J. (1975).
\newblock {Randomness and mathematical proof}.
\newblock {\em Scientific American\/}~{\em 232\/}(5), 47--53.

\bibitem[\protect\citeauthoryear{Choi and Fara}{Choi and
  Fara}{2021}]{sep-dispositions}
Choi, S. and M.~Fara (2021).
\newblock {Dispositions}.
\newblock In E.~N. Zalta (Ed.), {\em The {Stanford} Encyclopedia of
  Philosophy\/} ({S}pring 2021 ed.). Metaphysics Research Lab, Stanford
  University.

\bibitem[\protect\citeauthoryear{Cohnitz and Rossberg}{Cohnitz and
  Rossberg}{2024}]{sep-goodman}
Cohnitz, D. and M.~Rossberg (2024).
\newblock {Nelson Goodman}.
\newblock In E.~N. Zalta and U.~Nodelman (Eds.), {\em The {Stanford}
  Encyclopedia of Philosophy\/} ({S}pring 2024 ed.). Metaphysics Research Lab,
  Stanford University.

\bibitem[\protect\citeauthoryear{Crupi}{Crupi}{2021}]{sep-confirmation}
Crupi, V. (2021).
\newblock {Confirmation}.
\newblock In E.~N. Zalta (Ed.), {\em The {Stanford} Encyclopedia of
  Philosophy\/} ({S}pring 2021 ed.). Metaphysics Research Lab, Stanford
  University.

\bibitem[\protect\citeauthoryear{Dawid, Hartmann, and Sprenger}{Dawid
  et~al.}{2015}]{Dawid2015-DAWTNA}
Dawid, R., S.~Hartmann, and J.~Sprenger (2015).
\newblock {The No Alternatives Argument}.
\newblock {\em British Journal for the Philosophy of Science\/}~{\em 66\/}(1),
  213--234.

\bibitem[\protect\citeauthoryear{De~Benedetto and Luchetti}{De~Benedetto and
  Luchetti}{2024}]{de2024theory}
De~Benedetto, M. and M.~Luchetti (2024).
\newblock {Theory choice as niche construction: the feedback loop between
  scientific theories and epistemic values}.
\newblock {\em Philosophy of Science\/}~{\em 91\/}(3), 741--758.

\bibitem[\protect\citeauthoryear{Douglas}{Douglas}{2013}]{douglas2013value}
Douglas, H. (2013).
\newblock {The value of cognitive values}.
\newblock {\em Philosophy of science\/}~{\em 80\/}(5), 796--806.

\bibitem[\protect\citeauthoryear{Douven and G{\"a}rdenfors}{Douven and
  G{\"a}rdenfors}{2020}]{douven2020natural}
Douven, I. and P.~G{\"a}rdenfors (2020).
\newblock {What are natural concepts? A design perspective}.
\newblock {\em Mind \& Language\/}~{\em 35\/}(3), 313--334.

\bibitem[\protect\citeauthoryear{Duhem}{Duhem}{1954}]{Duhem1954}
Duhem, P. M.~M. (1954).
\newblock {\em {The Aim and Structure of Physical Theory}}.
\newblock Princeton: Princeton University Press.

\bibitem[\protect\citeauthoryear{Elgin}{Elgin}{1997}]{Elgin1997goodman}
Elgin, C.~Z. (Ed.) (1997).
\newblock {\em {The Philosophy of Nelson Goodman: Selected Essays}}.
\newblock New York: Garland.

\bibitem[\protect\citeauthoryear{Feigl}{Feigl}{1970}]{Feigl}
Feigl, H. (1970).
\newblock {The ``Orthodox'' View of Theories: Remarks in Defense as well as
  Critique}.
\newblock In M.~Radner and S.~Winokur (Eds.), {\em Minnesota Studies in the
  Philosophy of Science}, Volume~4, pp.\  3--16. University of Minnesota Press.

\bibitem[\protect\citeauthoryear{Feynman, Leighton, and Sands}{Feynman
  et~al.}{1963}]{Feynman1963}
Feynman, R.~P., R.~B. Leighton, and M.~L. Sands (1963).
\newblock {\em {The Feynman lectures on physics; New millennium ed.}}
\newblock New York, NY: Addison-Wesley Pub. Co.

\bibitem[\protect\citeauthoryear{Fletcher}{Fletcher}{2016}]{fletcher2016similarity}
Fletcher, S.~C. (2016).
\newblock {Similarity, topology, and physical significance in relativity
  theory}.
\newblock {\em The British Journal for the Philosophy of Science\/}~{\em
  67\/}(2), 365--389.

\bibitem[\protect\citeauthoryear{Fletcher}{Fletcher}{2024}]{fletcher2024alleged}
Fletcher, S.~C. (2024).
\newblock {On the Alleged Incommensurability of Newtonian and Relativistic
  Mass}.
\newblock {\em Erkenntnis\/}~{\em 90}, 3567--3588.

\bibitem[\protect\citeauthoryear{Fortunato, Bergstrom, Börner, Evans, Helbing,
  Milojević, Petersen, Radicchi, Sinatra, Uzzi, Vespignani, Waltman, Wang, and
  Barabási}{Fortunato et~al.}{2018}]{fortunato2018scisci}
Fortunato, S., C.~T. Bergstrom, K.~Börner, J.~A. Evans, D.~Helbing,
  S.~Milojević, A.~M. Petersen, F.~Radicchi, R.~Sinatra, B.~Uzzi,
  A.~Vespignani, L.~Waltman, D.~Wang, and A.-L. Barabási (2018).
\newblock Science of science.
\newblock {\em Science\/}~{\em 359\/}(6379), eaao0185.

\bibitem[\protect\citeauthoryear{G\"{a}rdenfors}{G\"{a}rdenfors}{1990}]{Gardenfors_1990}
G\"{a}rdenfors, P. (1990).
\newblock {Induction, Conceptual Spaces and AI}.
\newblock {\em Philosophy of Science\/}~{\em 57\/}(1), 78–95.

\bibitem[\protect\citeauthoryear{G{\"a}rdenfors}{G{\"a}rdenfors}{2000}]{Gardenfors_2000}
G{\"a}rdenfors, P. (2000).
\newblock {\em {Conceptual spaces: the geometry of thought}}.
\newblock A Bradford book. MIT Press.

\bibitem[\protect\citeauthoryear{G{\"a}rdenfors}{G{\"a}rdenfors}{2019}]{gardenfors2019reply_conde}
G{\"a}rdenfors, P. (2019).
\newblock {Convexity Is an Empirical Law in the Theory of Conceptual Spaces:
  Reply to Hern{\'a}ndez-Conde}.
\newblock In M.~Kaipainen, F.~Zenker, A.~Hautamäki, and P.~Gärdenfors (Eds.),
  {\em Conceptual Spaces: Elaborations and Applications}, Volume 405, pp.\ ~77.
  Springer.

\bibitem[\protect\citeauthoryear{G\"{a}rdenfors and Stephens}{G\"{a}rdenfors
  and Stephens}{2017}]{GardenforsStephens2017}
G\"{a}rdenfors, P. and A.~Stephens (2017).
\newblock {Induction and Knowledge-What}.
\newblock {\em European Journal for Philosophy of Science\/}~{\em 8\/}(3),
  1--21.

\bibitem[\protect\citeauthoryear{Glymour}{Glymour}{2013}]{glymour2013theoretical}
Glymour, C. (2013).
\newblock {Theoretical equivalence and the semantic view of theories}.
\newblock {\em Philosophy of science\/}~{\em 80\/}(2), 286--297.

\bibitem[\protect\citeauthoryear{Godfrey-Smith}{Godfrey-Smith}{2021}]{Godfreysmith2}
Godfrey-Smith, P. (2021).
\newblock {\em {Theory and Reality}\/} (2nd ed.).
\newblock Chicago: The University of Chicago Press.

\bibitem[\protect\citeauthoryear{Goodman}{Goodman}{1946}]{goodman1946confirmation}
Goodman, N. (1946).
\newblock {A Query on Confirmation}.
\newblock {\em Journal of Philosophy\/}~{\em 43}, 383--385.

\bibitem[\protect\citeauthoryear{Goodman}{Goodman}{1955}]{grue}
Goodman, N. (1955).
\newblock {\em {Fact, Fiction, and Forecast}\/} (2nd ed.).
\newblock Cambridge, MA: Harvard University Press.

\bibitem[\protect\citeauthoryear{Goodman}{Goodman}{1983}]{grue83}
Goodman, N. (1983).
\newblock {\em {Fact, Fiction, and Forecast}\/} (4th ed.).
\newblock Cambridge, MA: Harvard University Press.

\bibitem[\protect\citeauthoryear{Goodman}{Goodman}{2008}]{goodman2008pvalue}
Goodman, S. (2008).
\newblock {A Dirty Dozen: Twelve P-Value Misconceptions}.
\newblock {\em Seminars in Hematology\/}~{\em 45\/}(3), 135--140.
\newblock Interpretation of Quantitative Research.

\bibitem[\protect\citeauthoryear{Halvorson}{Halvorson}{2013}]{halvorson2013semantic}
Halvorson, H. (2013).
\newblock {The semantic view, if plausible, is syntactic}.
\newblock {\em Philosophy of science\/}~{\em 80\/}(3), 475--478.

\bibitem[\protect\citeauthoryear{Halvorson}{Halvorson}{2019}]{halvorson2019logic}
Halvorson, H. (2019).
\newblock {\em {The logic in philosophy of science}}.
\newblock Cambridge University Press Cambridge.

\bibitem[\protect\citeauthoryear{Hempel}{Hempel}{1943}]{hempel1943purely}
Hempel, C.~G. (1943).
\newblock {A purely syntactical definition of confirmation}.
\newblock {\em The Journal of Symbolic Logic\/}~{\em 8\/}(4), 122--143.

\bibitem[\protect\citeauthoryear{Henderson}{Henderson}{2024}]{Henderson-SEP}
Henderson, L. (2024).
\newblock {The Problem of Induction}.
\newblock In E.~N. Zalta (Ed.), {\em The {Stanford} Encyclopedia of
  Philosophy\/} ({W}inter 2024 ed.). Metaphysics Research Lab, Stanford
  University.

\bibitem[\protect\citeauthoryear{Hern\'{a}ndez{-}Conde}{Hern\'{a}ndez{-}Conde}{2017}]{HernandezC2017}
Hern\'{a}ndez{-}Conde, J.~V. (2017).
\newblock {A Case Against Convexity in Conceptual Spaces}.
\newblock {\em Synthese\/}~{\em 194\/}(10), 4011--4037.

\bibitem[\protect\citeauthoryear{Howson and Urbach}{Howson and
  Urbach}{1993}]{Bayesian}
Howson, C. and P.~Urbach (1993).
\newblock {\em {Scientific Reasoning: The Bayesian Approach}\/} (2nd ed.).
\newblock La Salle IL: Open Court Publishing Company.

\bibitem[\protect\citeauthoryear{Hume}{Hume}{1739}]{Hume}
Hume, D. (1739).
\newblock {\em {A Treatise of Human Nature}}.
\newblock Oxford: Oxford University Press.

\bibitem[\protect\citeauthoryear{Kelly}{Kelly}{2007}]{Kelly-efficiency}
Kelly, K.~T. (2007).
\newblock {Ockham’s Razor, Empirical Complexity, and Truth-finding
  Efficiency}.
\newblock {\em Theoretical Computer Science\/}~{\em 383}, 270--289.

\bibitem[\protect\citeauthoryear{Khamsi and Kirk}{Khamsi and
  Kirk}{2001}]{KhamsiKirk2001MetricSpaces}
Khamsi, M. and W.~Kirk (2001).
\newblock {\em An Introduction to Metric Spaces and Fixed Point Theory}.
\newblock John Wiley \& Sons, Ltd.

\bibitem[\protect\citeauthoryear{Kitcher}{Kitcher}{1989}]{Kitcher1989}
Kitcher, P. (1989).
\newblock {Explanatary Unification and the Causal Structure of the World}.
\newblock In {\em Scientific Explanation}, Volume~8, pp.\  410--505. University
  of Minnesota Press.

\bibitem[\protect\citeauthoryear{Kolmogorov}{Kolmogorov}{1965}]{Kolmogorov}
Kolmogorov, A.~N. (1965).
\newblock {Three Approaches to the Quantitative Definition of Information}.
\newblock {\em Problems Inform. Transmission\/}~{\em 1}, 1--7.

\bibitem[\protect\citeauthoryear{Kraj{\i}cek}{Kraj{\i}cek}{2004}]{krajicek2004proof}
Kraj{\i}cek, J. (2004).
\newblock {Proof complexity}.
\newblock In {\em European congress of mathematics (ECM), Stockholm, Sweden},
  pp.\  221--231.

\bibitem[\protect\citeauthoryear{Kuhn}{Kuhn}{1977}]{KuhnOVTC}
Kuhn, T.~S. (1977).
\newblock {Objectivity, Value, and Theory Choice}.
\newblock In {\em {The Essential Tension}}. Chicago: Chicago University Press.

\bibitem[\protect\citeauthoryear{Leitgeb}{Leitgeb}{2007}]{leitgeb2007newQA}
Leitgeb, H. (2007).
\newblock {A new analysis of quasianalysis}.
\newblock {\em Journal of Philosophical Logic\/}~{\em 36}, 181--226.

\bibitem[\protect\citeauthoryear{Leitgeb}{Leitgeb}{2024}]{Leitgeb2024-VC}
Leitgeb, H. (2024).
\newblock {Vindicating the Verifiability Criterion}.
\newblock {\em Philosophical Studies\/}~{\em 181\/}(1), 223--245.

\bibitem[\protect\citeauthoryear{McEldowney}{McEldowney}{2020}]{mceldowney2020morita}
McEldowney, P.~A. (2020).
\newblock {On Morita equivalence and interpretability}.
\newblock {\em The Review of Symbolic Logic\/}~{\em 13\/}(2), 388--415.

\bibitem[\protect\citeauthoryear{Mormann}{Mormann}{1995}]{mormann1995incompatible}
Mormann, T. (1995).
\newblock {Incompatible empirically equivalent theories: A structural
  explication}.
\newblock {\em Synthese\/}~{\em 103}, 203--249.

\bibitem[\protect\citeauthoryear{NIST}{NIST}{2019}]{meterSI}
NIST (2019).
\newblock {SI definition of Meter}.
\newblock \url{https://www.nist.gov/si-redefinition/meter}.

\bibitem[\protect\citeauthoryear{Norton}{Norton}{2021}]{norton2021material}
Norton, J.~D. (2021).
\newblock {\em The material theory of induction}.
\newblock University of Calgary Press.

\bibitem[\protect\citeauthoryear{Oberheim and Hoyningen-Huene}{Oberheim and
  Hoyningen-Huene}{2025}]{sep-incommensurability}
Oberheim, E. and P.~Hoyningen-Huene (2025).
\newblock {The Incommensurability of Scientific Theories}.
\newblock In E.~N. Zalta and U.~Nodelman (Eds.), {\em The Stanford Encyclopedia
  of Philosophy\/} (Spring 2025 ed.). Stanford University.

\bibitem[\protect\citeauthoryear{Okasha}{Okasha}{2007}]{okasha2007jackson}
Okasha, S. (2007).
\newblock What does goodman's ‘grue’problem really show?
\newblock {\em Philosophical Papers\/}~{\em 36\/}(3), 483--502.

\bibitem[\protect\citeauthoryear{Piattelli{-}Palmarini}{Piattelli{-}Palmarini}{1980}]{Piattelli1980}
Piattelli{-}Palmarini, M. (1980).
\newblock {\em {Language and Learning: The Debate Between Jean Piaget and Noam
  Chomsky}}.
\newblock Harvard University Press.

\bibitem[\protect\citeauthoryear{Pigliucci}{Pigliucci}{2021}]{demarcation-IEP}
Pigliucci, M. (2021).
\newblock {Pseudoscience and the Demarcation Problem}.
\newblock {\em The Internet Encyclopedia of Philosophy\/}~{\em ISSN-2161-0002},
  1.

\bibitem[\protect\citeauthoryear{Popper}{Popper}{1959}]{Popper}
Popper, K. (1959).
\newblock {\em {The Logic of Scientific Discovery}}.
\newblock New York: Basic Books.

\bibitem[\protect\citeauthoryear{Pulte}{Pulte}{1998}]{pulte1998jacobi}
Pulte, H. (1998).
\newblock {Jacobi's criticism of Lagrange: the changing role of mathematics in
  the foundations of classical mechanics}.
\newblock {\em Historia Mathematica\/}~{\em 25\/}(2), 154--184.

\bibitem[\protect\citeauthoryear{Putnam}{Putnam}{1962}]{PutnamWTAN}
Putnam, H. (1962).
\newblock {What Theories are Not}.
\newblock In E.~Nagel, P.~Suppes, and A.~Tarski (Eds.), {\em Logic, Methodology
  and Philosophy of Science}, pp.\  240--252. Stanford, Cal.: Stanford
  University Press.

\bibitem[\protect\citeauthoryear{Quine}{Quine}{1950}]{Quine2Dogs}
Quine, W. v.~O. (1950).
\newblock {Two Dogmas of Empiricism}.
\newblock {\em The Philosophical Review\/}~{\em 60}, 20--43.

\bibitem[\protect\citeauthoryear{Quine}{Quine}{1969}]{QuineOntoRel}
Quine, W. v.~O. (1969).
\newblock {\em {Ontological Relativity and Other Essays}}.
\newblock New York: Columbia University Press.

\bibitem[\protect\citeauthoryear{Quine}{Quine}{1975}]{quine1975empirically}
Quine, W. v.~O. (1975).
\newblock {On Empirically Equivalent Systems of the World}.
\newblock {\em Erkenntnis\/}~{\em 9}, 313.

\bibitem[\protect\citeauthoryear{Quine}{Quine}{1991}]{Quine2DogsRetro}
Quine, W. v.~O. (1991).
\newblock {Two Dogmas in Retrospect}.
\newblock {\em Canadian Journal of Philosophy\/}~{\em 21\/}(3), 265--274.

\bibitem[\protect\citeauthoryear{Roche and Sober}{Roche and
  Sober}{2023}]{roche2023purely}
Roche, W. and E.~Sober (2023).
\newblock {Purely probabilistic measures of explanatory power: A critique}.
\newblock {\em Philosophy of Science\/}~{\em 90\/}(1), 129--149.

\bibitem[\protect\citeauthoryear{Schindler}{Schindler}{2018}]{schindler2018theoretical}
Schindler, S. (2018).
\newblock {\em {Theoretical virtues in science: Uncovering reality through
  theory}}.
\newblock Cambridge University Press.

\bibitem[\protect\citeauthoryear{Scholz}{Scholz}{2024}]{Scholz2024CC}
Scholz, S. (2024).
\newblock {Conceptual Spaces: A Solution to Goodman's New Riddle of Induction?}
\newblock {\em Philosophia\/}~{\em 52\/}(4), 915--934.

\bibitem[\protect\citeauthoryear{Schurz}{Schurz}{2015}]{schurz2015ostensive}
Schurz, G. (2015).
\newblock {Ostensive learnability as a test criterion for theory-neutral
  observation concepts}.
\newblock {\em Journal for General Philosophy of Science\/}~{\em 46\/}(1),
  139--153.

\bibitem[\protect\citeauthoryear{Schwarz}{Schwarz}{1978}]{Schwarz78}
Schwarz, G. (1978).
\newblock {Estimating the Dimension of a Model}.
\newblock {\em Annals of Statistics\/}~{\em 4}, 461--464.

\bibitem[\protect\citeauthoryear{Scorzato}{Scorzato}{2013}]{Scorzato}
Scorzato, L. (2013).
\newblock {On the role of simplicity in science}.
\newblock {\em Synthese\/}~{\em 190}, 2867--2895.

\bibitem[\protect\citeauthoryear{Scorzato}{Scorzato}{2015}]{Scorzato-furbetto}
Scorzato, L. (2015).
\newblock {Science and Illusions}.
\newblock preprint: philsci-archive.pitt.edu/15570.

\bibitem[\protect\citeauthoryear{Scorzato}{Scorzato}{2016}]{Scorzato-silfs14}
Scorzato, L. (2016).
\newblock {A simple model of scientific progress}.
\newblock In L.~Felline, F.~Paoli, and E.~Rossanese (Eds.), {\em New
  Developments in Logic and Philosophy of Science}, Volume~3 of {\em SILFS}.
  College Publications.

\bibitem[\protect\citeauthoryear{Scorzato}{Scorzato}{2024}]{scorzato2024reliability}
Scorzato, L. (2024).
\newblock {Reliability and Interpretability in Science and Deep Learning}.
\newblock {\em Minds and Machines\/}~{\em 34\/}(3), 27.

\bibitem[\protect\citeauthoryear{Scorzato}{Scorzato}{2026}]{scorzato2026oxphos}
Scorzato, L. (2026).
\newblock {The oxidative phosphorylation controversy in the light of epistemic
  complexity}.
\newblock preprint: philsci-archive.pitt.edu/28769.

\bibitem[\protect\citeauthoryear{Sober}{Sober}{2015}]{sober2015ockham}
Sober, E. (2015).
\newblock {\em Ockham's razors: a user's manual}.
\newblock Cambridge University Press.

\bibitem[\protect\citeauthoryear{Sprenger and Hartmann}{Sprenger and
  Hartmann}{2019}]{SprengerHartmann-bayesian}
Sprenger, J. and S.~Hartmann (2019).
\newblock {\em {Bayesian Philosophy of Science: Variations on a Theme by the
  Reverend Thomas Bayes}}.
\newblock Oxford and New York: Oxford University Press.

\bibitem[\protect\citeauthoryear{Stalker}{Stalker}{1994}]{Stalker1994grue}
Stalker, D.~F. (Ed.) (1994).
\newblock {\em {Grue!: The New Riddle of Induction}}.
\newblock Chicago and La Salle, IL: Open Court.

\bibitem[\protect\citeauthoryear{Stanford}{Stanford}{2021}]{sep-scientific-underdetermination}
Stanford, K. (2021).
\newblock {Underdetermination of Scientific Theory}.
\newblock In E.~N. Zalta (Ed.), {\em The {Stanford} Encyclopedia of
  Philosophy\/} ({W}inter 2021 ed.). Metaphysics Research Lab, Stanford
  University.

\bibitem[\protect\citeauthoryear{Starr}{Starr}{2022}]{sep-counterfactuals}
Starr, W. (2022).
\newblock {Counterfactuals}.
\newblock In E.~N. Zalta and U.~Nodelman (Eds.), {\em The {Stanford}
  Encyclopedia of Philosophy\/} ({W}inter 2022 ed.). Metaphysics Research Lab,
  Stanford University.

\bibitem[\protect\citeauthoryear{Str\"{o}s{s}ner}{Str\"{o}s{s}ner}{2022}]{Strossner2022}
Str\"{o}s{s}ner, C. (2022).
\newblock {Criteria for Naturalness in Conceptual Spaces}.
\newblock {\em Synthese\/}~{\em 200\/}(2), 1--36.

\bibitem[\protect\citeauthoryear{Suppes}{Suppes}{1967}]{SuppesWST}
Suppes, P. (1967).
\newblock {What is a Scientific Theory}.
\newblock In S.~Morgenbesser (Ed.), {\em Philosophy o Science Today}, pp.\
  55--67. Basic Books.

\bibitem[\protect\citeauthoryear{Teller}{Teller}{1969}]{Teller1969projections}
Teller, P. (1969).
\newblock {Goodman's Theory of Projection}.
\newblock {\em British Journal for the Philosophy of Science\/}~{\em 20\/}(3),
  219--238.

\bibitem[\protect\citeauthoryear{van Fraassen}{van
  Fraassen}{1980}]{FraassenImage}
van Fraassen, B. (1980).
\newblock {\em {The Scientific Image}}.
\newblock Oxford, UK: Oxford University Press.

\bibitem[\protect\citeauthoryear{van Fraassen}{van Fraassen}{2014}]{van2014one}
van Fraassen, B. (2014).
\newblock {One or two gentle remarks about Hans Halvorson’s critique of the
  semantic view}.
\newblock {\em Philosophy of Science\/}~{\em 81\/}(2), 276--283.

\bibitem[\protect\citeauthoryear{Visser}{Visser}{1991}]{Visser1991}
Visser, A. (1991).
\newblock {The Formalization of Interpretability}.
\newblock {\em Studia Logica\/}~{\em 50\/}(1), 81--105.

\bibitem[\protect\citeauthoryear{Visser}{Visser}{2004}]{visser2004categories}
Visser, A. (2004).
\newblock {Categories of theories and interpretations}.
\newblock {\em Logic Group Preprint Series\/}~{\em 228}, 1--64.

\bibitem[\protect\citeauthoryear{Votsis}{Votsis}{2016}]{Votsis2016}
Votsis, I. (2016).
\newblock {Philosophy of Science and Information}.
\newblock In L.~Floridi (Ed.), {\em The Routledge Handbook of Philosophy of
  Information}. Routledge.

\bibitem[\protect\citeauthoryear{Wikipedia}{Wikipedia}{2025}]{wiki:Bielefeld_conspiracy}
Wikipedia (2025).
\newblock {Bielefeld conspiracy} --- {W}ikipedia{,} the free encyclopedia.
\newblock \url{https://en.wikipedia.org/wiki/Bielefeld_conspiracy}.
\newblock [Online; accessed 05-May-2025].

\bibitem[\protect\citeauthoryear{Williamson and Stanley}{Williamson and
  Stanley}{2001}]{williamson2001knowing}
Williamson, T. and J.~Stanley (2001).
\newblock Knowing how.
\newblock {\em Journal of Philosophy\/}~{\em 98\/}(8), 411--444.

\bibitem[\protect\citeauthoryear{Woodward}{Woodward}{2005}]{woodward2005making}
Woodward, J. (2005).
\newblock {\em {Making things happen: A theory of causal explanation}}.
\newblock Oxford university press.

\bibitem[\protect\citeauthoryear{Woodward and Ross}{Woodward and
  Ross}{2025}]{sep-scientific-explanation-20th}
Woodward, J. and L.~Ross (2025).
\newblock {20th Century Theories of Scientific Explanation}.
\newblock In E.~N. Zalta and U.~Nodelman (Eds.), {\em The {Stanford}
  Encyclopedia of Philosophy\/} ({W}inter 2025 ed.). Metaphysics Research Lab,
  Stanford University.

\bibitem[\protect\citeauthoryear{Zenil}{Zenil}{2020}]{Zenil2020review}
Zenil, H. (2020).
\newblock {A Review of Methods for Estimating Algorithmic Complexity: Options,
  Challenges, and New Directions}.
\newblock {\em Entropy\/}~{\em 22\/}(6), 1--28.

\end{thebibliography}

\end{document}